\documentclass[epj]{svjour}
\usepackage{epsfig}
\usepackage{amsmath}
\usepackage{amssymb}
\usepackage{graphicx}
\usepackage{ulem} 
\usepackage[usenames]{color}
\usepackage{bm}
\usepackage{epsf}

\newcommand{\be}{\begin{equation}}
\newcommand{\ee}{\end{equation}}
\newcommand{\ba}{\begin{eqnarray}}
\newcommand{\ea}{\end{eqnarray}}
\newcommand{\nn}{\nonumber}

\newcommand{\non}{\nonumber\\}

\newcommand{\Imass}{M_{Y}}
\newcommand{\Fmass}{M_{\Lambda^{*}}}

\newcommand{\pini}{p_{i}}
\newcommand{\pfin}{P}

\newcommand{\Amp}{{\cal M}}

\newcommand{\Pslash}{{P\hspace{-6pt}/ \hspace{3pt}}}

\newcommand{\intq}{\int \frac{d^{4}q}{(2\pi)^{4}}}

\begin{document}


\title{Helicity Amplitudes of the \boldmath$\Lambda(1670)$ and two \boldmath$\Lambda(1405)$ as dynamically generated resonances}
\author{M.~D\"oring\inst{1}\and D.~Jido\inst{2}\and E.~Oset\inst{3}}
\institute{
Institut f\"ur Kernphysik and J\"ulich Center for Hadron Physics, Forschungszentrum J\"ulich, D-52425 J\"ulich, Germany \thanks{\email{m.doering@fz-juelich.de}} 
\and Yukawa Institute for Theoretical Physics, Kyoto University, Kyoto, 606-8502, Japan \thanks{\email{jido@yukawa.kyoto-u.ac.jp}}
\and Departamento de F\'{\i}sica Te\'orica and IFIC, Centro Mixto Universidad de Valencia-CSIC, Institutos de Investigaci\'on de Paterna, Aptdo. 22085, 46071 Valencia, Spain \thanks{\email{oset@ific.uv.es}}
}

\date{ FZJ-IKP-TH-2010-4, YITP-10-7}

\abstract{
We determine the helicity amplitudes $A_{1/2}$ and radiative decay widths 
in the transition $\Lambda(1670) \to \gamma Y$ $(Y=\Lambda$ or $\Sigma^{0})$. 
The $\Lambda(1670)$ is treated as a dynamically generated resonance in meson-baryon 
chiral dynamics. We obtain the radiative decay widths of the $\Lambda(1670)$ to 
$\gamma \Lambda$ as $3\pm 2$ keV and to $\gamma \Sigma^{0}$ as $120\pm 50$ keV. 
Also, the $Q^{2}$ dependence of the helicity amplitudes $A_{1/2}$ is calculated. 
We find that the $K\Xi$ component in the $\Lambda(1670)$ structure, mainly responsible 
for the dynamical generation of this resonance, is also responsible for the significant 
suppression of the decay ratio $\Gamma_{\gamma\Lambda}/\Gamma_{\gamma\Sigma^{0}}$.
A measurement of the ratio would, thus, provide direct access to the nature of the 
$\Lambda(1670)$. To compare the result for the $\Lambda(1670)$, we calculate the 
helicity amplitudes $A_{1/2}$ for the two states of the $\Lambda(1405)$. Also, the 
analytic continuation of Feynman parameterized integrals of more complicated loop 
amplitudes to the complex plane is developed which allows for an internally consistent 
evaluation of $A_{1/2}$.
}


\maketitle


\section{Introduction}
Recent interest in the structure of baryon resonances has developed in the last decade since the experimental pro\-gress in measurements of hadron scattering and photon-induced hadronic  reactions has brought very precise and wide energy-range data.  In forthcoming experiments at J-PARC, intense kaon beams will enable us to investigate the properties of baryon resonances with strangeness.  For the theoretical study of baryon resonances, one of the conventional  descriptions is the constituent quark  model~\cite{Isgur:1978xj,Capstick:1986bm,Glozman:1995fu,Glozman:1996wq,Capstick:2000qj,Loring:2001kv,Merten:2002nz,Furuichi:2003eh,VanCauteren:2003hn,VanCauteren:2005sm}, in which the symmetries of quarks play a major role in the description of the resonance structure. For baryon  resonances decaying to mesons and a baryon under strong interaction, the hadron dynamics is also important to understand the structure. Such hadron dynamics can be implemented  by solving coupled channel scattering equations  in an isobar model~\cite{Drechsel:1998hk},  with phenomenological Hamiltonians~\cite{Penner:2002ma,Gasparyan:2003fp,Matsuyama:2006rp,Durand:2008es,Doring:2009yv}, or with chiral dynamics~\cite{Kaiser:1995eg,kaiser,angels,joseulf,bennhold,Jido:2003cb,nieves,carmen,hyodo,lutz,sarkar,Borasoy:2006sr,Oller:2006jw}.

The $\Lambda(1670)$ resonance is of special interest, because it has the same quantum numbers as the $\Lambda(1405)$,  $I=0$, $S=-1$ and $J^{p}=(1/2)^{-}$. While the $\Lambda(1405)$ appeared dynamically generated in Refs.~\cite{angels,Jido:2003cb} in the chiral unitary framework, it had been found later~\cite{bennhold} that within the very same scheme, also the $\Lambda(1670)$ resonance appears naturally, without the need to introduce explicit pole terms. In the studies of Refs.~\cite{Borasoy:2006sr,Oller:2006jw}, theoretical errors have been provided to the $I=0$, $S=-1$ scattering by using also the next-to-leading order chiral meson-baryon interaction.

Recently it has turned out in an analysis based on  the chiral unitary approach~\cite{Hyodo:2008xr} that  the $\Lambda(1405)$ is dominantly described by the meson-baryon  components, as suggested  since the late 50's~\cite{Dalitz:1959dn,Dalitz:1960du}, while to describe the $N(1535)$ some components other than meson and baryon are necessary.  Nevertheless, for the helicity amplitude of the $N(1535)$, the meson-baryon components give major contributions~\cite{Jido:2007sm,Jido:2008fr,Doring:2009qr}.

The investigation of the meson baryon components can be the touchstone to resolve the structure of baryon resonances; for this, the electromagnetic properties provide independent probes of the internal structure of the baryons. Thus, in this study we investigate the helicity amplitudes $A_{1/2}$  of the $\Lambda(1670)$ resonance. 

In Ref.~\cite{mishasourav}, the decays $\Lambda^*\to\gamma\Lambda$ and $\Lambda^*\to\gamma\Sigma^0$ of the $\Lambda(1520)$  were studied. It was concluded that there is strong meson-baryon dynamics, but that also a small genuine component is necessary. In the present case of the $\Lambda(1670)$, both decay modes are again possible, which further motivates this study in order to get more insight into the nature of this resonance.

In particular, it will turn out that the suppression of the ratio $\Gamma_{\gamma\Lambda}/\Gamma_{\gamma\Sigma^{0}}$ is closely tied to the presence of the $K\Xi$ channel, which by itself is mainly responsible for the dynamical generation of the $\Lambda(1670)$. Thus, the experimental determination of both radiative decay widths and their ratio would provide a test for the nature of the $\Lambda(1670)$.

In the present chiral unitary formulation, {\it i.e.}\ in the absence of genuine pole terms, the photo-excitation of the hyperon to the $\Lambda(1670)$ is expressed through the photon coupling to the constituent mesons and baryons~\cite{Jido:2007sm,mishasourav,Nacher:1999ni,Jido:2002yz,mishasolo,mishageng,Sekihara:2008qk}.  We use the model for the $\Lambda(1670)$ proposed  in Ref.~\cite{bennhold}, in which the couplings of the $\Lambda(1670)$ to the meson-baryon state have already been determined (see table 1 of Ref.~\cite{Ramos:2003wt} for values of the couplings in isospin basis). Thus, because the photon couplings to the mesons and baryons are given by the chiral Lagrangian, there are no free parameters and   the helicity amplitudes of the $\Lambda(1670)$ are pure predictions.  Photon couplings and gauge invariance in chiral unitary amplitudes were discussed in Refs.~\cite{Nacher:1999ni,Borasoy:2005zg,Borasoy:2007ku,Jido:2007sm,Sekihara:2008qk}. 

The helicity amplitude of the $\Lambda(1670)$ in a quark model was calculated in Ref.~\cite{VanCauteren:2005sm}, and we shall compare and discuss the results obtained with the two approaches [cf. Sec.~\ref{sec:resana}]. At the same time we evaluate the helicity amplitudes for the two $\Lambda(1405)$ states reported in Ref.~\cite{Jido:2003cb}, complementing the work done in Ref.~\cite{mishageng} where only the radiative decay widths were reported.

In studies of resonance properties using coupled channels  one works with real energies and generates amplitudes.  These amplitudes are directly comparable with experiment, but they have both resonance and non-resonant background  contributions. To resolve this entanglement and determine  the resonance properties  one can go to  the complex energy plane by making an analytic continuation  in the theoretical approach and evaluate the resonance properties  from the residues at the poles on the second Riemann sheet~\cite{Jido:2002yz,Sekihara:2008qk,Doring:2009bi}. The amplitudes obtained purely by the resonance contribution are relevant for comparison  with those obtained in other models. In Sec.~\ref{sec:newana}, we develop the analytic continuation of Feynman parameterized integrals with three propagators, as it appears in the one-loop amplitude of the $\Lambda^*MB$ vertex with the photon attached to the loop.  This allows for a consistent evaluation of the $\Lambda^*\to\gamma Y$ transition, with all quantities evaluated at the pole position of the $\Lambda(1670)$. 

However, in some experimental extractions of helicity amplitudes, many times unitary isobar models are used, in which $A_{1/2}$ appears as an effective, real coupling constant~\cite{Drechsel:2007if,Aznauryan:2009mx}. The connection to the residue formalism is not clear. We will thus also use another prescription in which the photon loop is evaluated at the real part of the pole position. This has been found to allow for a closer comparison to these experimentally extracted values~\cite{Jido:2002yz}.

Thus, following the calculation of the helicity amplitudes of the $N(1535)$  in Ref.~\cite{Jido:2007sm}, we first evaluate the helicity amplitudes of the $\Lambda(1670)$ and $\Lambda(1405)$ at the energies of the real parts of the pole positions and will then compare to the case of analytic continuation. As the $\Lambda(1670)$ and the two $\Lambda(1405)$ are quite narrow, both approaches are expected to give similar results.


\section{Formulation}

\begin{figure}
\begin{center}
\includegraphics[width=0.3\textwidth]{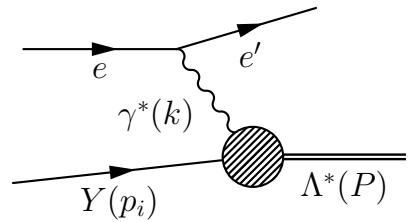}
  \caption{Kinematics of the electroproduction of the $\Lambda(1670)$. \label{fig1}}
\end{center}
\end{figure}

In this section, we explain the formalism to calculate the helicity  amplitude of a dynamically generated baryon resonance  in meson-baryon scattering. This method was developed  in Ref.~\cite{Jido:2007sm}.


\subsection{Helicity amplitude}
We consider the helicity amplitude $A_{1/2}$ of the $\Lambda(1670)$ resonance  $(\Lambda^{*})$ in the $\gamma^{*} Y \to \Lambda^{*}$ transition  with a virtual photon $\gamma^{*}$ and a hyperon $Y=\Lambda$ or $\Sigma^{0}$.  The helicity amplitude is defined in terms of the transition electric current $J_{\mu}$  of  the hyperon $Y$ ($J^{p}=(1/2)^{+}$) to  the $\Lambda^{*}$ with $J^{p}=(1/2)^{-}$ by
\begin{equation}
   A_{1/2}^{Y} = \sqrt \frac{2\pi \alpha}{q_{R}} \frac{1}{e} 
   \langle \Lambda^{*}, J_{z}=\frac{1}{2} | \epsilon^{(+)}_{\mu} J^{\mu} | Y, S_{z}=- \frac{1}{2} \rangle 
\end{equation}
with the fine structure constant $\alpha = e^{2}/4\pi$,  the energy equivalent to that of a real photon  $q_{R} =(W^{2}-\Imass^{2})/(2W)$ and the photon-hyperon center-of-mass energy $W\equiv \sqrt{\pfin^{2}}$.  The kinematic variables are defined in Fig.~\ref{fig1}. The polarization vector of the photon, $\epsilon_{\mu}^{(+)}$, is given in the  center-of-mass frame by
\begin{equation}
   \epsilon^{\pm}_{\mu} = \frac{1}{\sqrt 2} (0,\mp 1, -i ,0) \ ,
\end{equation}
where we take the momenta $\vec k$ and $\vec \pini$ along the $z$ axis. 

Lorentz invariance requires that  the general expression of the transition current  $J^{\mu}$  is given in the relativistic formulation by the following three Lorentz scalar amplitudes~\cite{Jido:2007sm}: 
\begin{equation}
   J^{\mu} = (\Amp_{1} \gamma^{\mu} + \Amp_{2} P^{\mu} + \Amp_{3} k^{\mu})\gamma_{5} . \label{eq:Jmu}
\end{equation}
Gauge invariance $  k \cdot J = 0$ reduces the number of independent amplitudes, $\Amp_{i}$ with the constraint 
\begin{equation}
   (\Fmass+\Imass) \Amp_{1} + k\cdot P \Amp_{2} + k^{2} \Amp_{3} = 0  \label{eq:iden} \ .
\end{equation}
In the calculations of the helicity amplitude, we take the resonance as an elementary particle and look at decay modes like in Fig.~\ref{fig:loops}. An alternative way of dealing with it is to look at the $\gamma\Lambda\to\bar KN$ amplitude and look at the pole of the $\Lambda(1670)$. We shall come back to this point later on. Thus, we have
\begin{equation}
   (\Pslash - \Fmass) u_{f} (\pfin)  = 0 \label{eq:EOMforNs}
\end{equation}
where $u_{f}(\pfin)$ is the $\Lambda^{*}$ Dirac spinor and  $\Fmass$ denotes the real part of the $\Lambda^{*}$ pole mass. 

In the rest frame of the $\Lambda^{*}$, the transition current (\ref{eq:Jmu})  can be written equivalently as
\begin{eqnarray}
  J^{\mu} &=&\sqrt{\frac{E_{i}+M_{Y}}{2M_{Y}}}\left[ \Amp_{1} \sigma^{\mu} 
  \right. \nn \\ && \left.
  + \left( \frac{\Amp_{1}}{(E_{i}+M_{Y})W} + \frac{\Amp_{2}}{E_{i}+M_{Y}} \right) P^{\mu} \sigma \cdot k
  \right. \nn \\ && \left.
  + \frac{\Amp_{3}}{E_{i}+M_{Y}} k^{\mu} \sigma \cdot k \right] 
  \label{eq:norecal} \\ 
  &\equiv& \Amp_{1}^{\rm NR} \sigma^{\mu} + \Amp^{\rm NR}_{2} P^{\mu} \sigma\cdot k + \Amp^{\rm NR}_{3} k^{\mu} \sigma \cdot k  \label{eq:nonreladecomp}
\end{eqnarray}
where $\sigma^{\mu} = (0, \vec \sigma)$ with the Pauli matrix $\sigma^{i}$  for the hyperon spin space and $P^{\mu}=(W , \vec 0)$ in the $\Lambda^{*}$ rest frame.  In Eq.~(\ref{eq:nonreladecomp}), we have defined the nonrelativistic amplitude   $\Amp^{\rm NR}_{i}$. In Eq.~(\ref{eq:norecal}), we have used the Dirac spinor for the initial hyperon normalized by
\begin{equation}
   u_{Y}(\pini) =\sqrt{\frac{E_{i}+\Imass}{2\Imass}}\left(
     \begin{array}{c} 1 \\ \frac{\vec \sigma \cdot \vec \pini}{E_{i} + \Imass} \end{array}
     \right ) \chi 
\end{equation}
and for the $\Lambda^{*}$ spinor $\bar u_{\Lambda^{*}} (P) = \chi^{\dagger} (1,0)$  in the $\Lambda^{*}$ rest frame. The gauge invariance condition  for the nonrelativistic amplitudes reads
\begin{equation}
  \Amp_{1}^{\rm NR}  + \Amp^{\rm NR}_{2} k\cdot P + \Amp^{\rm NR}_{3} k^{2}  = 0. \label{eq:gauseinvNR}
\end{equation}
With the transition current (\ref{eq:nonreladecomp}), the helicity amplitude, $A_{1/2}$, in the rest frame of the $\Lambda^{*}$ resonance, is written in terms of the amplitudes, $\Amp_{2}^{\rm NR}$ and $\Amp_{3}^{\rm NR}$,
as
\begin{equation}
   A_{1/2}^{Y} =  \sqrt \frac{2\pi \alpha}{q_{R}} 
   \frac{1}{e}
   \sqrt{2} \left( k\cdot P \Amp^{\rm NR}_{2} + k^{2} \Amp^{\rm NR}_{3}\right) \ .
   \label{eq:AampNR}
\end{equation}

The radiative decay width of $\Lambda^{*}$ to the hyperon $Y$,  $\Gamma_{\gamma}$, can be calculated with the helicity amplitude  $A_{1/2}$ at the real photon point by
\begin{equation}
   \Gamma_{\gamma Y} = \int d \Phi_{2} \left| \sqrt{E_{\gamma}} A_{1/2}^{Y} 
   \right|^{2}
\end{equation}
where $\Phi_{2}$ is the two-body phase space of the photon and hyperon  and defined as
\begin{equation}
   d\Phi_{2} = (2\pi)^{4} \delta^{(4)}(P-p_{\gamma}-p_{Y}) 
   \frac{d^{3}p_{\gamma}}{2E_{\gamma}(2\pi)^{3}}
  \frac{2M_{Y}}{2E_{Y}}\frac{d^{3}p_{Y}}{(2\pi)^{3}} \ .
\end{equation}
Performing the integral, one obtains the radiative decay width
\begin{equation}
   \Gamma_{\gamma Y} = \frac{q_{R}^{2}}{\pi} \frac{\Imass}{\Fmass} | A_{1/2}^{Y} |^{2} \label{eq:decaywidth}
\end{equation}


\subsection{Evaluation of the transition amplitude}
In the present approach, the $\Lambda^{*}$ is dynamically generated in meson-baryon scattering.  For the description of the $\Lambda(1670)$, the lowest lying octet mesons and baryons are the fundamental constituents  and interact with each other based on chiral dynamics.  The $\Lambda^{*}$ resonance is expressed as a pole of the scattering amplitude $T_{ij}(W)$ in the complex energy plane.  The details of the description of the amplitude $T_{ij}$ for the $\Lambda(1670)$ are given in Appendix~\ref{app:L1670}. 

\begin{figure}
\includegraphics[width=0.47\textwidth]{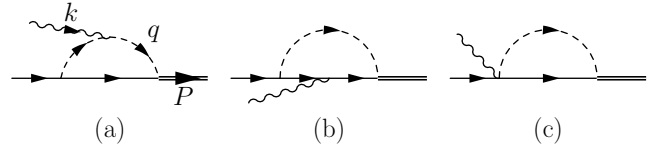}  
\caption{Feynman diagrams for the phototransition to the $\Lambda^{*}$. The solid, dashed, wavy and double lines denote octet baryons, mesons, photon and $\Lambda^{*}$, respectively. \label{fig:FeynmanDiagram}}
  \label{fig:loops}
\end{figure}

In this work, we follow the method developed in Ref.~\cite{Jido:2007sm}, which is  similar to the approaches in Refs.~\cite{mishasourav,mishasolo,mishageng} for real photons.  In the meson-baryon picture, the phototransition of the hyperon $Y$ to  the $\Lambda^{*}$ resonance is formulated through the photon couplings  to the constituent mesons and baryons of the $\Lambda^*$.  The Feynman diagrams for the transition amplitudes are shown in  Fig.~\ref{fig:FeynmanDiagram}. In the loops, all possible mesons and baryons contribute. We sum up all the contributions of the meson-baryon channels  to the transition amplitudes.   In the Feynman diagrams, (a) and (b) are the meson and baryon pole terms in which the photon attaches to the meson and baryon in the loop,  respectively. Diagram (c) contains the Kroll-Ruderman coupling  of the contact interaction for the photon, meson and baryon.  
For the electric transition there are no diagrams in which the photon attaches to the external baryons, since we consider the phototransition of the neutral baryons, while the magnetic transition is possible for the neutral baryon, but the contributions of the magnetic coupling were found to be small~\cite{Jido:2007sm}.
In Ref.~\cite{Jido:2007sm}, we have shown gauge invariance of the amplitude obtained  by summing up these diagrams in the relativistic calculation. 

This was done assuming the resonance as an elementary particle. An alternative approach would be to consider the scattering process $\gamma \Lambda \to\bar K N$ and look at the pole of the $\Lambda(1670)$ resonance.  If one does not stick to the resonance pole but looks into a wider range of energies the issue of gauge invariance becomes more involved and it is addressed in Ref.~\cite{Borasoy:2007ku}.

In Ref.~\cite{Borasoy:2007ku}, the fully gauge invariant amplitude of kaon photo- and electroproduction has been constructed. There, full gauge invariance is guaranteed by coupling the photon to all possible vertices and propagators of the hadronic rescattering series. In this case, the processes from Fig.~\ref{fig:loops} comprise not all of these processes and are also evaluated in a different framework of the on-shell approximation of the scattering potential. In any case, the processes considered in this study, i.e. the photon coupling to the first meson-baryon loop of the rescattering series, are numerically the most important ones around the resonance energy. They form a gauge invariant subclass of diagrams as shown in Refs.~\cite{Doring:2009uc,Doring:2009qr}: the process with the photon coupling directly to the Weinberg-Tomozawa interaction can be shrinked to a generalized contact current keeping gauge invariance of the resulting amplitude. Then, the resulting amplitude contains only the transverse part of the $u$- and $t$-channel contributions in the FSI loop contribution.  The generalized contact current results in the dressed Kroll-Ruderman term, i.e. with physical $MBB$ coupling constant.

The basic interactions of the mesons and baryons are given by the chiral Lagrangian: 
\begin{eqnarray}
{\cal L}_{MBB} &=& 
 - \frac{D}{\sqrt 2 f} \, {\rm Tr} \left[ \bar B \gamma_{\mu} \gamma_{5} \{\partial^{\mu}\Phi,B\} \right]
 \nonumber \\ && 
 - \frac{F}{\sqrt 2 f}\, {\rm Tr} \left[ \bar B \gamma_{\mu} \gamma_{5} [\partial^{\mu}\Phi ,B] \right] \label{eq:MBcoup}
\end{eqnarray}
with the meson and baryon fields, $\Phi$ and $B$, defined by
\begin{eqnarray}
   \Phi &=& 
    \left(
   \begin{array}{ccc}
       \frac{1}{\sqrt{2}} \pi^0 + \frac{1}{\sqrt{6}} \eta & \pi^+ & K^+ \\
       \pi^- & -\frac{1}{\sqrt{2}} \pi^0 + \frac{1}{\sqrt{6}} \eta  & K^0\\
       K^- & \bar K^0 & - \frac{2}{\sqrt{6}} \eta
   \end{array}
   \right)
\\
   B &=& 
   \left(
   \begin{array}{ccc}
       \frac{1}{\sqrt{2}} \Sigma^0 + \frac{1}{\sqrt{6}} \Lambda &
       \Sigma^+ & p \\
       \Sigma^- & -\frac{1}{\sqrt{2}} \Sigma^0 + 
       \frac{1}{\sqrt{6}} \Lambda  & n\\
       \Xi^- & \Xi^0 &- \frac{2}{\sqrt{6}} \Lambda
   \end{array}
   \right) \ .
\end{eqnarray}
The $MBB$ couplings are obtained from these Lagrangian as $g_{A}^{i}/(2f)$ where $g_{A}^{i}$ is the axial vector coupling listed in Table~\ref{tab:gA} and $f$ is the meson decay constant. In this work, we use $f=1.123 f_{\pi}$  with $f_{\pi}=93$ MeV~\cite{bennhold} for all channels, and the values of $D$ and $F$ for the axial vector couplings are taken from Ref. \cite{Luty:1993gi} as
\begin{equation}
D=0.85 \pm 0.06 \ ,\ \ \ \ \ F=0.52 \pm 0.04 \ .  \label{eq:DFvalues}
\end{equation}
The $D$ and $F$ values were fixed by the experimental data of  the hyperon axial vector couplings, neglecting higher-order corrections in chiral  perturbation theory. The couplings of photon to  mesons and  baryons are given by the gauge couplings:
\begin{eqnarray}
    {\cal L}_{\gamma B} &=& -e {\rm Tr}\left[ \bar B \gamma_{\mu} [Q_{\rm ch},B]\right] A^{\mu} \label{eq:LaggBB} \\
    {\cal L}_{\gamma M} &=& ie {\rm Tr}\left[ \partial_{\mu} \Phi [Q_{\rm ch},\Phi]\right] A^{\mu}
\end{eqnarray}
with the charge matrix $Q_{\rm ch}={\rm diag}(\frac{2}{3},-\frac{1}{3},-\frac{1}{3})$ and $e>0$.
By replacing the derivative acting on the meson fields $\partial_{\mu}\Phi$ with the covariant derivative $D_{\mu}\Phi = \partial_{\mu}\Phi + i e A_{\mu} [Q_{\rm ch},\Phi]$  in the interaction Lagrangian (\ref{eq:MBcoup}), we obtain the Kroll-Ruderman terms of the $\gamma MBB$ couplings, which are proportional to the meson charge $Q_{M}$.  For the couplings of the $\Lambda^{*}$ resonance to  the meson and baryon,  we take a Lorentz scalar form representing the $s$-wave nature:
\begin{equation}
   {\cal L}_{\Lambda^{*}M_{i}B_{i}} = g_{\Lambda^{*}}^{i} \bar \Lambda^{*} \Phi_{i} B_{i} \ .
   \label{effective}
\end{equation}
The coupling strengths are obtained by the residua of the scattering  amplitude at the $\Lambda(1670)$ pole position in the chiral unitary  approach and the values are given in Table \ref{tab:coupl1680}. Note that $g_i\neq 0$ for the pure isospin $I=1$ channels, which originates from the small isospin breaking due to the use of physical masses in the coupled channel scheme. For details of the residue determination, see Sec.~\ref{sec:newana}.

\begin{table*}
\caption{The axial vector coupling $g_{A}^{i}$ for each channel obtained by the flavor SU(3) symmetry. The values of $D$ and $F$ are given in Eq.~(\ref{eq:DFvalues}).
\label{tab:gA}}
\begin{center}
\begin{tabular}{cccccc}  
\hline
 channel &  
$\Lambda pK^{-} $ & $\Lambda n \bar K^{0}$ & $\Lambda \Lambda \pi^{0}$ & $\Lambda \Sigma^{0} \pi^{0}$  & $\Lambda \Lambda \eta$ \\
 $g_{A}^{i}$ & $-\frac{1}{\sqrt 3} (D + 3 F)  $ & $-\frac{1}{\sqrt 3} (D + 3 F) $ & $0$ & $\frac{2}{\sqrt 3} D$& $-\frac{2}{\sqrt 3} D$\\
 with Eq.~(\ref{eq:DFvalues}) & $-1.39$ & $-1.39$ &  $0$ & $ 0.98$ & $ -0.98$ \\
\hline
channel & $\Lambda \Sigma^{0} \eta$ &$\Lambda \Sigma^{-} \pi^{+} $& $\Lambda \Sigma^{+} \pi^{-}$ & $\Lambda \Xi^{-} K^{+}$ & $\Lambda \Xi^{0} K^{0}$ \\
 $g_{A}^{i}$ & $0$ & $\frac{2}{\sqrt 3} D$ & $\frac{2}{\sqrt 3} D$ & $\frac{1}{\sqrt 3} (-D + 3 F) $ & $ \frac{1}{\sqrt 3} (-D + 3 F) $ \\
  with Eq.~(\ref{eq:DFvalues}) & $ 0$ & $0.98$ &  $0.98$ & $ 0.41$ & $ 0.41$ \\
\hline\hline
 channel &  $\Sigma^{0} pK^{-} $ & $\Sigma^{0} n \bar K^{0}$ & $\Sigma^{0} \Lambda \pi^{0}$ & $\Sigma^{0} \Sigma^{0} \pi^{0}$  & $\Sigma^{0} \Lambda \eta$ \\
 $g_{A}^{i}$ &   $D-F$ & $-D+F $ & $\frac{1}{\sqrt 3} D $ & $0$ & $0$ \\
  with Eq.~(\ref{eq:DFvalues}) & $ 0.33$ & $-0.33$ &  $0.98$ & $ 0$ & $ 0$ \\
 \hline
 channel & $\Sigma^{0} \Sigma^{0} \eta$ &$\Sigma^{0} \Sigma^{-} \pi^{+} $& $\Sigma^{0} \Sigma^{+} \pi^{-}$ & $\Sigma^{0} \Xi^{-} K^{+}$ & $\Sigma^{0} \Xi^{0} K^{0}$ \\
$g_{A}^{i}$& $\frac{2}{\sqrt 3} D $ & $2F$ & $-2F$ & $D+F$ & $-D-F$ \\
  with Eq.~(\ref{eq:DFvalues}) & $ 0.98$ & $1.04$ &  $-1.04$ & $ 1.37$ & $ 1.37$ \\ \hline
\end{tabular}
\end{center}
\end{table*}

\begin{table*}
\caption{Complex coupling constants {$g_i$} of $\Lambda(1670)$ to the meson-baryon channels. These values are obtained as the residua of the meson-baryon scattering amplitude at the $\Lambda(1670)$ pole position $z=1680-20 i$ MeV in the chiral unitary model. See Appendix \ref{app:L1670} for the details.}
\begin{center}
\begin{tabular}{cccccc}
 \hline\hline
channel &   $pK^{-} $ & $n \bar K^{0}$ & $\Lambda \pi^{0}$ & $\Sigma^{0} \pi^{0}$  & $\Lambda \eta$ \\
$g_{\Lambda^{*}}^{i}$ &   $ -0.212 + 0.498 i $ & $-0.216 + 0.517 i $ & $ -0.008-0.013 i$ & $-0.003  + 0.153 i$ & $1.050-0.111i$ \\ 
$|g_{\Lambda^{*}}^{i}|$&   $0.541$ & $0.560$ & $0.015$ & $0.153$ & $1.056$ \\
 \hline
channel &$\Sigma^{0} \eta$ &$\Sigma^{-} \pi^{+} $& $\Sigma^{+} \pi^{-}$ & $\Xi^{-} K^{+}$ & $\Xi^{0} K^{0}$ \\
$g_{\Lambda^{*}}^{i}$ &  $ -0.008-0.012i$& $-0.007+ 0.141 i$ & $  0.000+0.164i$  & $2.429-0.103 i$ & $2.452-0.071i$\\
$|g_{\Lambda^{*}}^{i}|$&    $0.014$ & $0.142$ & $0.164$ & $2.431$ & $2.453$ \\
  \hline\hline
\end{tabular}
\end{center}
\label{tab:coupl1680}
\end{table*}

\begin{figure}[t]
\epsfxsize= 8.5cm
\begin{center}
  \epsfbox{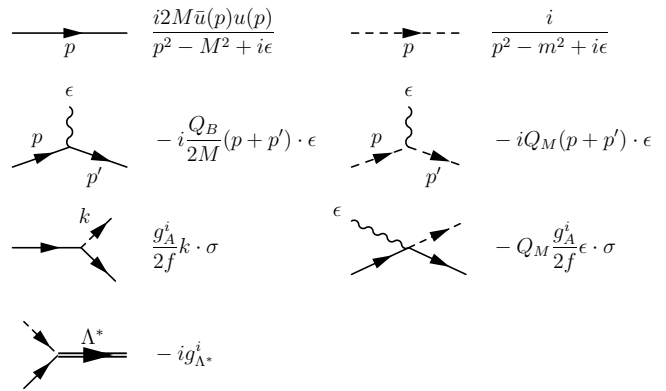}
\end{center}
  \caption{Nonrelativistic Feynman rules for the propagator and the elementary vertices. The solid, dashed, wavy and double lines denote octet baryons, mesons, photon and $\Lambda^{*}$, respectively. $M$ and $m$ denote the baryon and meson masses, respectively. $Q_{B}$ and $Q_{M}$ are the charges of the baryon and meson. \label{fig:FeynmanRule}}
\end{figure}

We calculate the transition amplitudes in the nonrelativistic formulation, since it has turned out that the nonrelativistic calculation is enough for the  low-energy transition amplitude for the $N(1535)$ in Ref.~\cite{Jido:2007sm}. For the transition amplitude of the hyperons to the $\Lambda(1670)$, the photon energy is comparable to the transition of the nucleon to the $N(1535)$. In the  nonrelativistic formulation, we take only the leading terms of the  $1/M$ expansion which are the diagrams (a) and (c) in Fig.\ref{fig:FeynmanDiagram}, as shown in Ref.~\cite{Jido:2007sm}.  Diagram (b) is of next-to-leading order due to the $1/M$ factor  in the $\gamma BB$ coupling. Thus we do not taken into account diagram (b) in our final result of the nonrelativistic calculation. Note that in pion photoproduction, the $\kappa/(2M)$ contribution and kinetic corrections can be large~\cite{Bernard:1992nc,Bernard:1994gm}. In order to see the influence of the anomalous magnetic moment $\kappa$, the corresponding loop contribution has been calculated in Ref.~\cite{Jido:2007sm}. It is of similar, small size as the convection current in the diagram with $\gamma BB$ coupling. See also the discussion in Sec.~\ref{sec:resana} where we consider these processes for the determination of the theoretical error.

The diagrams (a) and (c) can be decomposed in terms of the Lorentz structures given  in Eq.~(\ref{eq:nonreladecomp}). Among the three Lorentz invariant amplitudes,  the helicity amplitude can be expressed by  $\Amp_{2}^{\rm NR}$ and $\Amp_{3}^{\rm NR}$ as shown in  Eq.~(\ref{eq:AampNR}).  It is also found in Ref.~\cite{Jido:2007sm} that diagram (c) has only the  $\Amp_{1}^{\rm NR}$ term. Thus, we do not calculate the diagram (c).  In fact, the amplitudes $\Amp_{2}^{\rm NR}$ and $\Amp_{3}^{\rm NR}$ remain finite even with one loop integration, and $\Amp_{1}^{\rm NR}$ does have a divergence in the loop calculation, which cancels with divergences coming from the other diagrams thanks to gauge invariance. 

In Ref.~\cite{mishasolo} this has been confirmed for the case of a neutral external baryon by explicitly calculating the set of gauge invariant diagrams of Fig.~\ref{fig:loops}. The logarithmic divergences of the diagrams cancel in their sum. The sum equals the result obtained by using the condition of gauge invariance of Eq. (\ref{eq:gauseinvNR}), which is the method adopted in this study.

The Feynman rules for the nonrelativistic couplings are given in Fig.~\ref{fig:FeynmanRule}.  In the figure, $\epsilon^{\mu}$ denotes the photon polarization, and $\sigma$ is  the Lorentz covariant form of the spin matrix, $\sigma^{\mu} = (0,\vec \sigma)$.  $Q_{B}$ and $Q_{M}$ are the baryon and meson charges such that  $Q_{B}$ is $e>0$ for the proton with $e^{2}/(4\pi) = \alpha \simeq 1/137$.  $g_{A}^{i}$ stands for the axial vector coupling constants of the baryons to  the corresponding meson. The values of $g_{A}^{i}$ for each channel are  given in Table \ref{tab:gA}. 
$g_{\Lambda^{*}}^{i}$ is the coupling strength of the $\Lambda^{*}$ to  the meson-baryon channel $i$.  The values are listed in Table \ref{tab:coupl1680}.
For the baryon propagator, we use the covariant form  from Eq. (\ref{covaform}).  For the evaluation of diagram (b), we use  the $\gamma BB$ vertex obtained by a nonrelativistic reduction  of the interaction Lagrangian (\ref{eq:LaggBB}) as
\begin{eqnarray}
   -i Q_{B} \bar u \gamma \cdot \epsilon u &\rightarrow& - iQ_{B} \chi^{\dagger} \left[ \epsilon^{0} - \frac{\vec\epsilon \cdot (\vec p + \vec p^{\, \prime} )}{2M} \right] \chi   \label{eq:NRgBB1} \\
   &\simeq&  - iQ_{B} \chi^{\dagger} \left[  \frac{\epsilon \cdot (p + p^{\, \prime} )}{2M} \right ] \chi
   \label{eq:NRgBB2}
\end{eqnarray}
where we have used the fact that the baryon kinetic energies are small in the nonrelativistic kinematics, $p^{0} \simeq p^{0\prime} \simeq M$,  in the last expression. In Eqs.~(\ref{eq:NRgBB1}) and (\ref{eq:NRgBB2}) we have neglected the magnetic term that behaves like $(\vec \sigma \times \vec k) / 2M $ which has one power less in the loop variable.

The amplitude $-it=J\cdot \epsilon$ for the diagram (a) with channel $i$ is calculated by applying the Feynman rules shown in Fig.\ref{fig:FeynmanRule} as
\begin{eqnarray}
   -it_{a}^{i}  & = & 
   iQ_{M} A_{i}   \int \frac{d^{4} q}{(2\pi)^{4}} 
   \frac{  ( q - k) \cdot \sigma  \,   (2q-k)\cdot \epsilon}
   {(P-q)^{2} - M^{2}_{i}+ i\epsilon}\non
   &\times&\frac{1}{(q^{2} - m^{2}_{i} + i\epsilon)((q-k)^{2} - m^{2}_{i}+ i\epsilon )} 
\end{eqnarray}
with the coefficient $A_{i}$ defined by
\begin{equation}
 A_{i}= \frac{  g_{A}^{i} g_{\Lambda^{*}}^{i} M_{i}}{f }\ . \label{eq:defA}
\end{equation}
We use the Feynman parameterization of the integral
\begin{equation}
   \frac{1}{abc} = 2 \int^{1}_{0} dx \int^{x}_{0} dy \frac{1}{\left(a + (b-a) x + (c-b)y\right)^{3}} \ . \label{eq:FeynParaInt}
\end{equation}
and the integral variable $q^{\prime}$, such that $q=q^{\prime}+P(1-x) + ky$ and renaming $q^{\prime}$ as $q$, we eliminate the linear terms of $q$  in the denominator and obtain
\begin{eqnarray}
   -it_{a}^{i}   & = & 
 iQ_{M} A_{i} 2 \int^{1}_{0} dx \int^{x}_{0} dy 
   \int \frac{d^{4} q}{(2\pi)^{4}} 
   \frac{  ( q +(y-1) k) \cdot \sigma }
   { \left(q^{2} - S_{a}^{i} + i\epsilon \right)^{3}} \non
&\times&   \left(2q+(2y-1)k +2(1-x) P\right)\cdot \epsilon
   \ , \label{eq:diaaNR}
\end{eqnarray}
where we use $P \cdot \sigma = 0$ in the CM frame and $S_{a}^{i}$ is defined by
\begin{eqnarray}
   S_{a}^{i} &=&  2 P\cdot k (1-x)y - P^{2}x(1-x) -  k^{2} y (1-y)   \nn \\ &&
   + M^{2}_{i} (1-x) + m_{i}^{2}x \ . \label{eq:defSa}
\end{eqnarray}
In Eq.~(\ref{eq:diaaNR}), only even powers of $q$ give contribution after performing  the integration.  The $q^{\mu}q^{\nu}$ term in the numerator contributes to the $\Amp^{\rm NR}_{1}$  and is divergent.  The terms with 0th power of $q$ remain finite and contribute to the $\Amp^{\rm NR}_{2}$ and $\Amp^{\rm NR}_{3}$ amplitudes. Finally after performing the integration,  we get the $\Amp^{\rm NR}_{2}$ and $\Amp^{\rm NR}_{3}$ components  for the channel $i$ as
\begin{eqnarray}
    \Amp^{i \rm (NR)}_{2a} &=&    \frac{ Q_{M} A_{i}}{(4\pi)^{2}} 
    \int^{1}_{0} dx \int^{x}_{0} dy  \frac{2(y-1)(1-x)}{S_{a}^{i} - i\epsilon } \label{eq:amp2NR}
    \\
    \Amp^{i \rm (NR)}_{3a} &=&   \frac{ Q_{M} A_{i}}{(4\pi)^{2}}
    \int^{1}_{0} dx \int^{x}_{0} dy  \frac{(y-1)(2y-1)}{S_{a}^{i} - i\epsilon } \label{eq:amp3NR}
\end{eqnarray}
where we have used
\begin{eqnarray}
  \intq \frac{1}{(q^{2}-S)^{3}} &=& -\frac{i}{(4\pi)^{2}} \frac{1}{2}
  \left(\frac{1}{S}\right)\ .  \label{eq:LoopS3}
\end{eqnarray}

In a similar way we evaluate the finite contribution from diagram (b) which,  as we mentioned, is of order $1/M$ of the previous ones and we obtain
\begin{eqnarray}
    \Amp^{i \rm (NR)}_{2b} &=&   -\frac{ Q_{B} A_{i}}{(4\pi)^{2}} 
    \int^{1}_{0} dx \int^{x}_{0} dy  \frac{2y(1-x)}{S_{b}^{i}-i\epsilon}\non
    \Amp^{i \rm (NR)}_{3b} &=&   - \frac{ Q_{B} A_{i}}{(4\pi)^{2}}
    \int^{1}_{0} dx \int^{x}_{0} dy  \frac{y(2y-1)}{S_{b}^{i}-i \epsilon}
    \label{convection}
\end{eqnarray}
with
\begin{eqnarray}
   S_{b}^{i} &= & 2 P\cdot k (1-x)y - P^{2}x(1-x) -  k^{2} y (1-y) 
   \nn \\ && 
   + m^{2}_{i} (1-x) + M_{i}^{2}x \ . \label{eq:defSb}
\end{eqnarray}

Finally, the helicity amplitude is obtained by summing up all the channels and substituting the amplitudes (\ref{eq:amp2NR}) and (\ref{eq:amp3NR}) in Eq.~(\ref{eq:AampNR}). 

We also multiply the transition amplitudes obtained above by the electromagnetic
form factors of the mesons or baryons to which the photon couples, in order to
take into account the charge distribution of the constituent mesons and baryons
in the resonance. The $Q^2$ dependence of the helicity amplitude of  the
$\Lambda^{*}$ resonance, thus, stems from the form factors  of the meson and
baryons components and the intrinsic  $Q^2$ structure of the loops. For the
mesons and baryons form factors, we  take monopole form factors consistent with the values for the radii of the mesons:
\begin{equation}
F(Q^2) = \frac{\Lambda^2}{\Lambda^2+ Q^2} \label{eq:FormFactor}
\end{equation}
with
\begin{eqnarray}
\Lambda_\pi &=& 0.727\ {\rm [GeV]} \\
\Lambda_K &=& 0.828\ {\rm [GeV]} .
\end{eqnarray}
These values correspond to $\langle r^2 \rangle=0.44$ fm$^2$ and  $\langle r^2 \rangle=0.34$ fm$^2$ for the pion and the kaon,  respectively \cite{Amendolia:1986wj,Amendolia:1986ui,Oller:2000ug}. These values were used in Ref.~\cite{Oller:2000ug} in the study of the pion and kaon form factors using unitary chiral theory, where very good results were obtained for both form factors in a relatively large range of momentum transfers. Hence, we stick to these values here. For the baryon form factor, we take the same form as for the corresponding meson to keep  gauge invariance. Given the practically negligible effect of the baryon terms in the evaluation of the helicity amplitudes, the approximation made there has no practical consequences.


\section{Analytic continuation of helicity amplitudes}
\label{sec:newana}
As the expressions for the amplitudes $\cal{M}$ in Eqs.~(\ref{eq:amp2NR}), (\ref{eq:amp3NR}), and (\ref{convection}) show, the hadronic $MB\Lambda^*$ vertices are encoded in the values of the coupling constants $g_{\Lambda^*}$ as defined in Eq.~(\ref{eq:defA}) for the different meson-baryon channels $(MB)_i$. These coupling constants are obtained from the $\Lambda^*$ pole residues in the complex plane at $z=z_0$ (in the following, we use the variable $z\equiv \sqrt{P^2}$ to indicate that the energy can become complex). The residue is obtained in the expansion of the $MB\to MB$ scattering amplitude $T_{(2)}^{ij}$ from channel $i$ to channel $j$,
\ba
a_{-1}&\equiv&  g_{\Lambda^*}^i\,g_{\Lambda^*}^j,\non
T_{(2)}^{ij}&=&\frac{a_{-1}}{z-z_0}+a_0+a_1(z-z_0)+{\cal O}(z^2).
\label{pa}
\ea
The pole is situated on the second sheet of the scattering amplitude, $T_{(2)}$. The analytic continuation from the first sheet $T_{(1)}$ to $T_{(2)}$ is given by the continuation of the meson-baryon loop function $G$ and the hadronic vertices $V$ in the scattering equation (cf. Eq. (\ref{bse})) and has been discussed many times in the literature~\cite{Jido:2003cb}.

The second parts of the amplitudes $\cal{M}$ in Eqs.~(\ref{eq:amp2NR}), (\ref{eq:amp3NR}), and (\ref{convection}) are given by Feynman  parameterized loop functions. To obtain the resonance properties without non-resonant background contamination, the integrals should be evaluated at the resonance position $z=z_{0}$ on their second sheet as done in Refs.~\cite{Jido:2002yz,Sekihara:2008qk} for the magnetic moment and electromagnetic radii of $\Lambda(1405)$. For the helicity amplitudes, the necessary analytic continuation was not available until now, and in previous studies~\cite{Jido:2007sm,mishageng} the amplitude $\cal{M}$ has been calculated by evaluating  those loop functions at an energy of $z={\rm Re}\, z_0$ instead of $z=z_0$.  In this section, we show how to construct the analytic continuation of  Feynman parameterized integrals; in Sec.~\ref{sec:resana},  numerical results at $z={\rm Re}\, z_0$ and $z=z_0$ will be compared.


\subsection{Analytic continuation of Feynman parameterized integrals}
We concentrate here on the dominant meson pole term [cf. l.h.s. of Fig.~\ref{fig:loops}] associated with the amplitude given in Eqs.~(\ref{eq:amp2NR}) and (\ref{eq:amp3NR}). For real photons ($k^2=0$), the term of Eq. (\ref{eq:amp2NR}) alone already provides the leading order contribution. The amplitude can be rewritten as
\ba
&&\Amp^{\rm (NR)}_{(2a),\,j}(k^2=0) =    \frac{2 Q_{M} A_{j}}{(4\pi)^{2}} \,\tilde d_j, \quad
\tilde d_j =\int\limits_0^1 dx\int\limits_0^{1-x}dy \non
&&\times\frac{x(y-1)}{x[(x-1)z^2+y(z^2-M_e^2)+M_j^2]+(1-x)m_j^2-i\,\epsilon}\non
\label{dterma}
\ea
where $M_j\,(m_j; M_e)$ is the mass of the intermediate baryon (meson; external baryon, in this case $\Lambda$ or $\Sigma^0$). The loop depends on the total c.m. energy $z$. The expression in Eq. (\ref{dterma}) can be explicitly evaluated with the result given in Ref. \cite{mishasolo}.

For the continuation of $\tilde d_j$, one could use the method of contour deformation~\cite{Suzuki:2008rp,Doring:2009yv} for the integrals over Feynman parameters $x$ and $y$, but in the present case, an explicit evaluation of the imaginary along the right-hand cut, and thus a way to compensate this discontinuity, is feasible and allows for a straightforward solution: 

The imaginary part of $\tilde d_j$ above threshold $z>z_{\rm thres}=m_j+M_j$ is given by
\ba
i\,{\rm Im}\,\tilde d_j&=&i\pi\int\limits_0^1 dx\,\frac{a+1}{M_e^2-z^2}\, \theta(-a)\,\theta(a+1-x),\non
a&=&\frac{x[z^2(x-1)+M_j^2]+(1-x)m_j^2}{x(z^2-M_e^2)}.
\label{ima1}
\ea
For the continuation, it is necessary to explicitly evaluate the $\theta$ functions. It turns out that for all $z>m_j+M_j$, $a+1-x>0$, i.e. the second $\theta$ function can be omitted. From the first $\theta$ function, one obtains for the imaginary part at $z>m_j+M_j$ ($z$ real):
\ba
i\,{\rm Im}\,\tilde d_j&=&i\pi\int\limits_{x_-}^{x_+} dx\,\frac{a+1}{M_e^2-z^2},\quad
x_\pm=\frac{E_m\pm q_{cm}}{z}\non
\label{impa}
\ea
where 
\ba
E_m&=&\frac{z^2-M_j^2+m_j^2}{2z},\non
q_{cm}&=&\frac{\sqrt{(z^2-(m_j+M_j)^2)(z^2-(m_j-M_j)^2)}}{2z}\non
\label{limits}
\ea
are the on-shell meson energy and relative momentum. 

Eq. (\ref{impa}) explicitly shows the dependence of the integration limits on the total energy $z$. With this information, the analytic continuation is straightforward: The task is to analytically continue the function $\delta \tilde d_j$ where
\ba
\delta \tilde d_j=i\,{\rm Im}\,\tilde d_j\quad \text{for}\,z\geq m_j+M_j
\ea
and then add $\delta \tilde d_j$ twice to remove the discontinuity of the right-and cut. Note that $\delta \tilde d_j\neq i\,{\rm Im}\,\tilde d_j$ for complex $z$ (the function $i$~Im~$\tilde d_j$ is not analytic in $z$). 

In the complex plane, $\delta \tilde d_j$ is still given by Eq. (\ref{impa}), the only difference being that the integral becomes now a contour integral in the complex $x$ plane from complex $x_-$ to complex $x_+$.

For complex values of $z$ one has to be careful with the two Riemann sheets of the square root appearing in $q_{cm}$ of Eq. (\ref{limits}). Furthermore, for ${\rm Re}\,z$ sufficiently smaller than threshold, one has to deform the straight integration path from $x_-$ to $x_+$ because of the singularity at $x=0$ appearing in the expression for $a$ in Eq. (\ref{ima1}). The final result, valid in the whole complex plane, is then given by the contour integration
\ba
\delta\tilde d_j&=&i\pi\int\limits_\Gamma dx\,\frac{a+1}{M_e^2-z^2}\ ,\non
\Gamma&=&
\begin{cases}
\Gamma(x_1,x_2)&\text{if Re $x_1>0\,\land$ Re$\,x_2>0$}\\
\Gamma(x_1,x_c,x_2)&\text{else (choose $x_c=0.1$)}
\end{cases}\non
x_1&=&
\begin{cases}
x_-& \text{if Im $x_-\geq 0$}\\
x_+& \text{if Im $x_- < 0$}
\end{cases}\ ,\non
x_2&=&
\begin{cases}
x_+& \text{if Im $x_-\geq 0$}\\
x_-& \text{if Im $x_- < 0$}
\end{cases}\ .
\label{final2}
\ea
The arguments of $\Gamma$ are given by the edge-points of the piecewise straight integration path $\Gamma$. The contour deformation with $x_c>0$ (e.g. $x_c=0.1$) avoids the pole at $x=0$ and the redefinition $x_\pm\to x_{1,2}$ selects the correct Riemann sheet of the square root in $q_{cm}$ from Eq. (\ref{limits}).

\begin{figure}
\begin{center}
\includegraphics[width=0.47\textwidth]{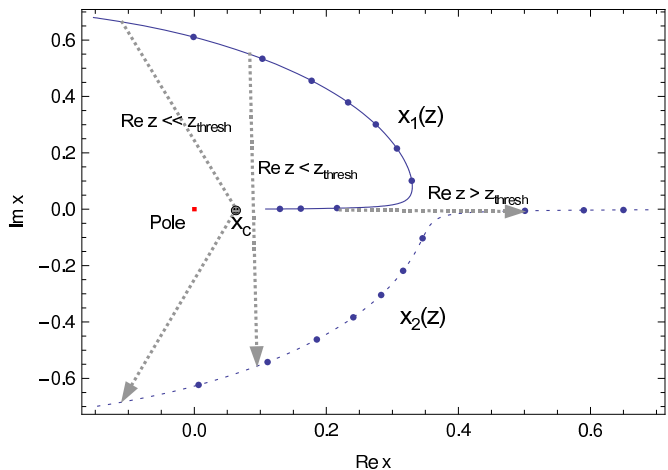}
\end{center}
\caption{(Color Online) Integration limits $x_{1,2}$ for the calculation of $\delta\tilde d_{\bar KN}$ of the $\bar KN$ loop with external $\Lambda$ according to Eq. (\ref{final2}). The solid (dashed) lines show $x_1\,(x_2)$ as a function of $z$ being varied from $1800-5\,i$ MeV to $700-5\,i$ MeV. The integration contours $\Gamma$ are indicated with the dotted arrows. The case Re $z>z_{\rm thresh}$ indicates the case of (almost) real $x_{1,2}$, i.e. $z$ (almost) real and above threshold. When Re $z$ becomes Re $z<z_{\rm thresh}$, the $x_\pm$ acquire an imaginary part. For some $z<<z_{\rm thresh}$, one has to slightly deform the integration contour [indicated as $\Gamma(x_1,x_c,x_2)$ in Eq. (\ref{final2})] to avoid the pole at $x=0$, as indicated in the figure.}
\label{fig:contour}
\end{figure}

In Fig. \ref{fig:contour}, Eq. (\ref{final2}) is illustrated for the example of a $\bar KN$ loop. The positions of the integration limits in the complex $x$ plane are shown as a function of the total energy $z$ being varied between $1800-5\,i$ MeV down to $700-5\,i$ MeV. For energies above threshold, the contour is just a normal integration along the real $x$ axis. Below threshold, the integration limits acquire an imaginary part, and for energies far below threshold, one has to slightly deform the integration contour to avoid the pole at $x=0$. 

Another question concerns the possibility of additional branch points in the complex plane for $\delta\tilde d_j$. In Ref. \cite{Doring:2009yv} it has been shown that branch points arise if an integration limit coincides with a pole of the integrand. However, in the present case, $x_{1,2}\neq 0$ for all complex energies $z$, and no additional branch points are induced from the pole at $x=0$. 

Having determined $\delta\tilde d_j$, the analytic continuation of $\tilde d_j$ is given by adding twice $\delta \tilde d_j$ (the discontinuity along the right-hand cut equals twice the imaginary part),
\be
\tilde d_j^{\,(2)}=\tilde d_j+2\,\delta\tilde d_j\ . 
\label{analcont}
\ee
Summarizing, $\tilde d_j\equiv \tilde d_j^{\,(1)}$ is the first sheet with a branch point at threshold, $z_{\rm thres}=m_j+M_j$, and $\tilde d_j^{\,(2)}$ is the second sheet. The branch cut where $\tilde d_j^{\,(1)}$ and $\tilde d_j^{\,(2)}$ are analytically connected, is given by the right-hand cut which runs from threshold to $+\infty$ along the real $z$ axis. This analytic structure is, thus, very similar to the meson-baryon loop function $G$ [cf. Eqs. (\ref{bse}), (\ref{eq:gpropdr})].

Note that $\tilde d_j^{\,(2)}$ has a pole at $z^2=M_{\rm e}^2$ [cf. Eq. (\ref{ima1})] which is, however, cancelled in the expression for $A_{1/2}$ through the $k\cdot P=(z^2-M_{\rm e}^2)/2$ term in Eq. (\ref{eq:AampNR}). 

\begin{figure}
\begin{center}
\includegraphics[width=0.47\textwidth]{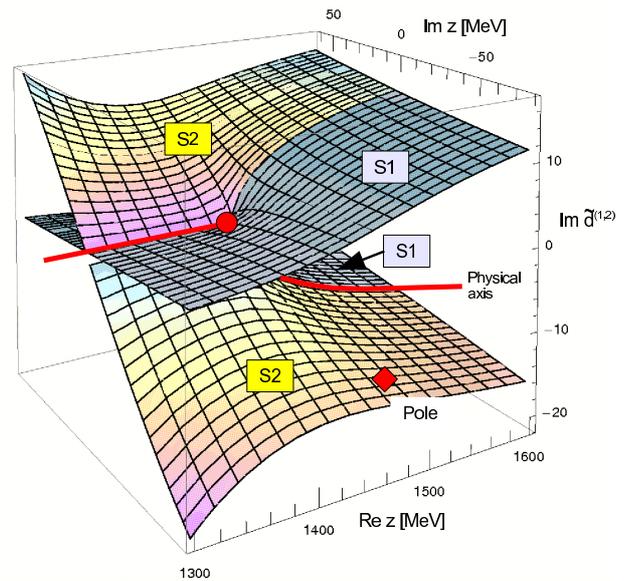}
\end{center}
\caption{(Color Online) Imaginary part of the two Riemann sheets $\tilde d_j^{(1)}\equiv \tilde d_j$ and $\tilde d_j^{(2)}$ of the $\bar K N$ meson pole loop [cf. Fig.~\ref{fig:loops} (a)]. The two sheets are indicated as S1 and S2, respectively. As indicated in the figure, the physical axis is connected to the first sheet S1 in the upper $z$ half plane, and to the second sheet S2 in the lower half plane. The right-hand cut is clearly visible along which the two Riemann surfaces are analytically connected. Also, the branch point at $z=m_{\bar K}+M_N$ is visible (circle). Resonance poles are on the second sheet S2 (diamond).}
\label{fig:analcont}
\end{figure}

In Fig. \ref{fig:analcont}, the first and second sheet $\tilde d_j^{(1)}\equiv \tilde d_j$ and $\tilde d_j^{\,(2)}$ of the $\bar K N$ meson pole loop are shown. The right-hand cut is clearly visible along which the two Riemann surfaces are analytically connected. Also, the branch point at $z=m_{\bar K}+M_N$ is visible.
The physical axis at $z+i\,\epsilon$ belongs to the first Riemann sheet S1 in the upper half plane. Thus, if one evaluates $A_{1/2}$ at a resonance pole in the lower half plane, one has to use the analytic continuation to evaluate S2. 

For the actual coupled channel calculation with channels $j$, we use the same selection of Riemann sheets as for the meson-baryon propagator $G$ itself~\cite{Jido:2003cb} [cf. Eqs. (\ref{bse}), (\ref{eq:gpropdr}); $G$ in this work serves only to calculate the hadronic coupling constants $g_{\Lambda^*}$ from Eq. (\ref{effective})]. It is constructed to choose the unphysical Riemann sheet most closely connected to the physical, real $z$ axis according to
\ba
\Amp^{\rm (NR)}_{(2a),\,j}=\frac{2 Q_{M} A_{j}}{(4\pi)^{2}}
\begin{cases}
\tilde d_j^{\,(2)}& \text{if Re $z \geq m_j+M_j$}\\
\tilde d_j^{\,(1)}& \text{else}
\end{cases}\ .
\non
\ea
This expression is valid for real photons ($k^2=0$); the extension to virtual photons is straightforward, and the analytic continuation for the amplitude associated with the baryon pole term, $\Amp^{i \rm (NR)}_{2b, 3b}$ from Eq. (\ref{convection}), can be constructed in analogy to the present case of the meson pole term.


\subsection{Proof of analyticity}
The proof of analyticity of Eq. (\ref{analcont}) follows Ref.~\cite{Doring:2009yv}, but is modified. Although we treat here only the case of a real photon, the argument is similar for virtual photons. Note that the form factor from Eq. (\ref{eq:FormFactor}) factorizes from the loop amplitude and does not induce new analytic structures in the $z$ plane for virtual photons.
\begin{figure}
\begin{center}
\includegraphics[width=0.4\textwidth]{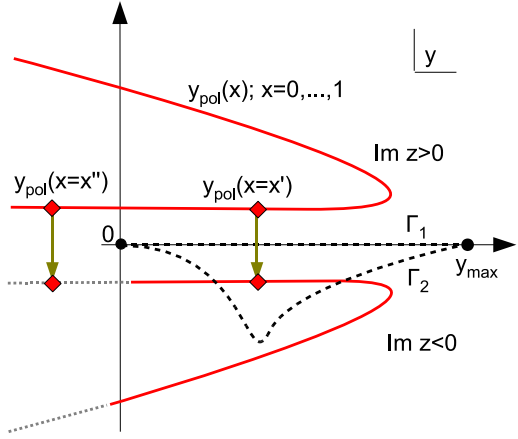}
\end{center}
\caption{(Color Online) Plane of the integration variable $y$ from Eq. (\ref{dterma}). The integration path of the first sheet of $\tilde d_j$ is along the path $\Gamma_1$ (dashed line). The position of the pole of the integrand depends on the other integration variable $x\in [0,1]$ and is indicated with the thick solid red line for a case Im $z>0$ (integrand pole in the upper $y$ half plane) and a case Im $z<0$ (integrand pole in the lower $y$ half plane). If Im $z$ changes from positive to negative, the pole position $y_{\rm pol}(x=x')$ crosses the integration contour $\Gamma_1$. Thus, to analytically continue along the right-hand cut, one has to deform the contour $\Gamma_1\to\Gamma_2$.}
\label{fig:proof}
\end{figure}
Consider the integration over $x$ and $y$ in the expression for $\tilde d_j$ from Eq. (\ref{dterma}). The plane of complex $y$ is shown in Fig. \ref{fig:proof} together with the integration contour $\Gamma_1$ from 0 to $y_{\rm max}=1-x$ (dashed line). The pole of the integrand is given by 
\be
y_{\rm pol}(x,z)=-a
\ee 
with $a$ from Eq. (\ref{ima1}). For Im $z>0$ (Im $z<0$), the pole lies in the upper (lower) $y$ half plane for all $x\in[0,1]$. The pole positions for these two cases are indicated in the figure with the red solid lines. Consider now a $z$ with Re $z>m_j+M_j$ and a small Im $z>0$. The pole position is indicated as $y_{\rm pol}(x=x')$ in the figure. As Im $z$ becomes smaller and finally negative, at the same time the pole position crosses the integration contour $\Gamma_1$. This induces the right hand cut of the first sheet of $\tilde d_j$, i.e. the discontinuity of $\tilde d_j$ from threshold to infinity along the real $z$ axis. To achieve the analytic continuation along the right-hand cut, the integration contour has to be deformed for negative Im $z$. This is indicated as $\Gamma_2$ in the figure. The result is then by construction analytically connected to $\tilde d_j$ in the upper $z$ half plane. 

Considering a fixed $x$ for the moment, the difference $\Delta \tilde d_j$ of the $y$ integration along $\Gamma_2$ (second sheet) minus the integration along $\Gamma_1$ (first sheet) is given by the residue, 
\ba
&&\Delta \tilde d_j=\int_{\Gamma_2}dy\,\cdots-\int_{\Gamma_1}dy\,\cdots=2\pi i\,\rm{Res}_{y=-a}\non
&&\times \big[\frac{x(y-1)}{x[(x-1)z^2+y(z^2-M_e^2)+M_j^2]+(1-x)m_j^2}\big]\non
&&=2\pi i\,\frac{a+1}{M_e^2-z^2}\ .
\label{diffres}
\ea
where the dots stand for the integrand of $\tilde d_j$. Providing the $x$ integration,
\be
\int_\Gamma dx\, \Delta \tilde d_j=\tilde d_j^{(2)}-\tilde d_j^{(1)}=2\,\delta\tilde d_j
\ee
which proves that $\tilde d_j^{(2)}$ from Eq. (\ref{analcont}) is indeed not only a continuous but also the analytic continuation of $\tilde d_j^{(1)}$ along the right-hand cut.

As for the limits of the $x$-integration from Eq. (\ref{final2}) becoming complex for complex $z$, note that in general integration limits have to be analytically continued in order the integral to be analytic (for illustration, consider the counterexample of the integral $\int_0^{|z|}dz=|z|$ over the analytic function 1, which is non-analytic for all $z\in \mathbb{C}$). 

In Fig. \ref{fig:proof}, note that if Im $z$ changes sign, there are values of $x$ for which the pole does not cross $\Gamma_1$. In the figure, this is indicated as $y_{\rm pol}(x=x'')$. In this case, the integration contour must not be deformed, and the corresponding $x$ values do not contribute to $\Delta \tilde d_j$. This fact is automatically taken care of by the explicit limits $x_{1,2}$ in the $x$ integration of Eq. (\ref{final2})~\footnote{Strictly speaking, the fact that the integration limits $x_{1,2}$ are complex for complex $z$, changes the picture of Fig. \ref{fig:proof} slightly, because the pole position depends on $x$, $y_{\rm pol}=-a(x,z)$. However, this does not change the proof presented here.}.   


\section{Results}

\label{sec:results}
\subsection{Results for the \boldmath{$\Lambda(1670)$}}
In this section, we discuss our numerical results for the  transition amplitude of the $\Lambda(1670)$ resonance.  For the calculation of the transition amplitude, we fix the center-of-mass energy of the $\Lambda(1670)$ at $W=1680$ MeV which is the real part of the $\Lambda(1670)$ pole position in the meson-baryon scattering amplitude. The influence of the analytic continuation developed in the previous section will be discussed in Sec.~\ref{sec:resana}.
\begin{figure}
\includegraphics[width=0.48\textwidth]{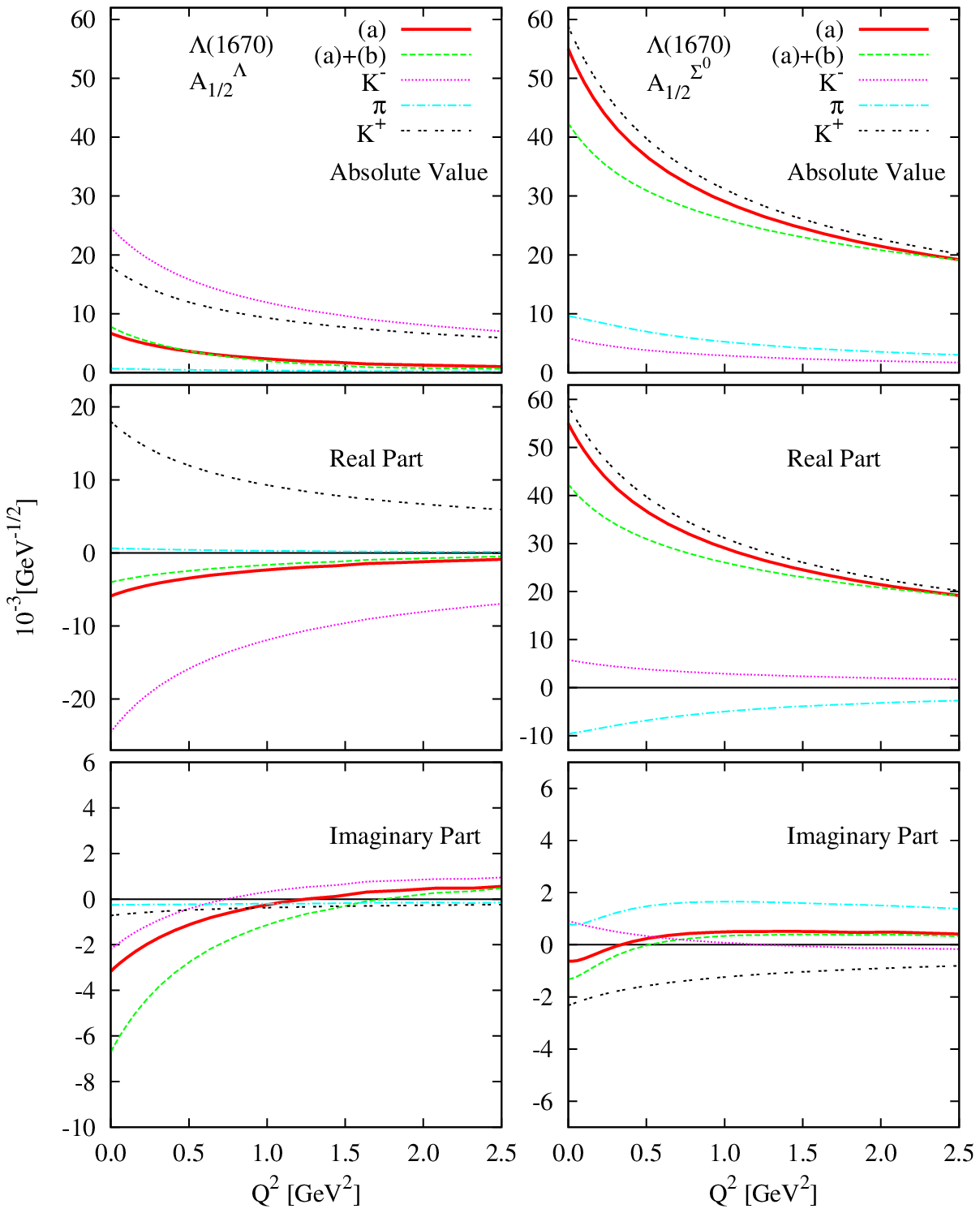} 
\caption{(Color Online) $A_{1/2}^{p}$ helicity amplitudes for the  $\gamma \Lambda \to\Lambda(1670)$ (left) and  $\gamma \Sigma^{0} \to\Lambda(1670)$ (right) transitions calculated as a function of $Q^{2}$ in unit of $10^{-3}$ GeV$^{-1/2}$.  The upper, middle  and lower panels are respectively the modulus, real parts and imaginary parts  of the amplitudes.  The solid lines show the calculation with diagram (a) (meson pole term).  The dashed line stands for the calculations of sum of diagrams (a) and (b).  The dotted, dot-dashed and two-dotted lines denote the $K^{-}$, pion and $K^{+}$  contributions, respectively.\label{fig:L1670A}  }
\end{figure}

\begin{table*}
\caption{Helicity amplitude at $Q^{2}=0$ in unit of $10^{-3}$ GeV$^{-1/2}$.
The indices (a) and (a)$+$(b) denote the calculations with diagram (a) 
(meson pole term)
and with sum of diagrams (a) and (b) (baryon pole term) in Fig.~\ref{fig:FeynmanDiagram}, and $K^{-}$, $\pi$ and $K^{+}$ stand for 
the contributions from the $K^{-}$, pion ($\pi^{+}$ and $\pi^{-}$) and 
$K^{+}$ pole terms, respectively. 
Both $\Lambda(1380)$ and $\Lambda(1426)$ are the resonance states
for the $\Lambda(1405)$, indicating the states at the pole positions of
$1390-66i$ MeV and $1426-16i$ MeV, respectively (see Appendix~\ref{app:L1670}).}
\label{tab:A1/2value}
\begin{center}
\begin{tabular}{c|cc|cc|cc}
    \hline \hline
    & \multicolumn{6}{c}{ $\gamma \Lambda \to  \Lambda^{*}$ } \\
    \hline
   & \multicolumn{2}{c}{ $\Lambda(1670)$} &  \multicolumn{2}{c}{ $\Lambda(1390)$} &
    \multicolumn{2}{c}{  $\Lambda(1426)$} \\
   \hline
   & $A_{1/2}$ &  $|A_{1/2}|$ & $A_{1/2}$ &  $|A_{1/2}|$ & $A_{1/2}$ &  $|A_{1/2}|$ \\
 (a)   & $  -5.9   -3.2 i$ & $   6.7$   & $ -23.8  + 40.3 i$ & $  46.8$  & $  76.6  + 29.5 i$ & $  82.1$\\
(a) $+$ (b) & $  -4.0   -6.7 i$ & $   7.8$ & $ -19.1  + 32.7 i$ & $  37.8$ & $  63.7 +  24.8 i$ & $  68.3$  \\
 $K^{-}$    & $ -24.5   -2.2 i$ & $  24.6$& $ -25.0  + 34.1 i$ & $  42.3$  & $  74.9 +  29.7 i$ & $  80.6$ \\
  $\pi$  & $   0.6   -0.3 i$ & $   0.7$ & $   2.1   + 5.3 i$ & $   5.7$ & $   1.4   -1.0 i$ & $   1.7$ \\
 $K^{+}$    & $  18.0   -0.7 i$ & $  18.0$ & $  -1.0  +  0.9 i$ & $   1.3$ &  $   0.3 +  0.8 i$ & $   0.9$	\\
   \hline\hline
    & \multicolumn{6}{c}{$\gamma \Sigma^{0} \to  \Lambda^{*}$ } \\
    \hline
   & \multicolumn{2}{c}{ $\Lambda(1670)$} &  \multicolumn{2}{c}{ $\Lambda(1390)$} &
    \multicolumn{2}{c}{  $\Lambda(1426)$} \\
    \hline
   & $A_{1/2}$ &  $|A_{1/2}|$ & $A_{1/2}$ &  $|A_{1/2}|$ & $A_{1/2}$ &  $|A_{1/2}|$ \\
 (a)     & $  55.0   -0.6 i$ & $  55.0$  & $-104.9  -35.9 i$ & $ 110.8$ & $ -54.7 +  38.7 i$ & $  67.0$ \\
 (a) $+$ (b)  & $  42.4   -1.3 i$ & $  42.4$ & $ -89.9  -41.2 i$ & $  98.9$ & $ -52.1 +  29.4 i$ & $  59.8$ \\
  $K^{-}$  & $   5.8  + 0.9 i$ & $   5.9$  & $   5.7   -7.7 i$ & $   9.6$  & $ -17.8   -7.1 i$ & $  19.1$ \\
   $\pi$  & $  -9.6   + 0.8 i$ & $   9.6$   & $-107.7  -30.9 i$ & $ 112.0$ & $ -37.8  + 43.4 i$ & $  57.6$  \\
   $K^{+}$   & $  58.8   -2.3 i$ & $  58.9$ 	&  $  -2.8  +  2.7 i$ & $   3.9$  & $   0.9  +  2.4 i$ & $   2.6$ 	\\
  \hline \hline
\end{tabular}
\end{center}
\end{table*}

In Table~\ref{tab:A1/2value} we list the values of the helicity  amplitudes  of $\Lambda(1670)$ at $Q^{2}=0$ for the real photon, and  in Fig.~\ref{fig:L1670A} we show the $Q^{2}$ dependence of the  helicity amplitudes for the $\gamma \Lambda \to\Lambda(1670)$  (left side) and $\gamma \Sigma^{0} \to\Lambda(1670)$ (right side) transition. In Fig.~\ref{fig:L1670A},  the upper, middle and lower panels are respectively the modulus,  real parts and imaginary parts of the amplitudes.  The solid lines are the results of the calculation with the diagram (a)  (meson pole term) shown in Fig.~\ref{fig:FeynmanDiagram}.  In the dashed line, we add the contribution of the diagram~(b) (baryon pole-term), which is of next-to-leading order in the $1/M$ expansion.  Comparing the solid and dashed lines, we find that the relativistic corrections are not important for the helicity amplitudes in these energies.  We also calculate the contributions to the helicity amplitudes from  the  $K^{-}$, pion ($\pi^{+}$ and $\pi^{-}$) and $K^{+}$ pole terms, which are denoted by the dotted, dot-dashed and two-dotted lines, respectively.

From Table~\ref{tab:A1/2value} and Fig.~\ref{fig:L1670A}, we find that the helicity amplitude in the  $\gamma \Lambda \to \Lambda(1670)$ transition is one order of magnitude smaller than that in the  $\gamma \Sigma^{0} \to \Lambda(1670)$ one. This is because, in the $\gamma\Lambda \to \Lambda(1670)$ transition, there is a large cancellation between the $K^{-}$ and $K^{+}$  pole terms, while in the $\gamma\Sigma^{0} \to \Lambda(1670)$ transition, the $K^{+}$ contribution is dominant.  The $K^{+}$ pole term dominance can be understood by  the $\Lambda(1670)$ coupling to the meson-baryon states. As seen in Table~\ref{tab:coupl1680}, the $\Lambda(1670)$ dominantly couples to $K^{+}\Xi^{-}$, while the coupling to $K^{-}p$  is sub-dominant and those to $\pi^{\pm}\Sigma^{\mp}$ are negligibly small among the states having charged meson and baryon.   In addition, according to the values of the $MBB$ coupling $g_{A}$ shown in Table~\ref{tab:gA}, the $\Sigma^{0} \Xi^{-} K^{+}$ coupling  is also large. Thus the $K^{+}$ pole term is dominant in  the $\gamma\Sigma^{0} \to \Lambda(1670)$ transition.  In the $\gamma\Lambda \to \Lambda(1670)$ transition, although the $\Lambda(1670)$ has a large coupling to $K^{+}\Xi^{-}$ and a small coupling to $K^{-}p$, since the $\Lambda p K^{-}$ coupling is about three times larger than the $\Lambda \Xi^{-} K^{+}$ coupling, the contributions of the  $K^{-}$ and $K^{+}$ pole terms are comparable in magnitude  and cancel each other.  It is also interesting to interpret the larger helicity amplitude  of the $\gamma\Sigma^{0} \to \Lambda(1670)$ transition with respect to that of $\gamma\Lambda\to \Lambda(1670)$  in terms of isospin symmetry. Since the $\Lambda(1670)$ has $I=0$, the isosinglet component of the photon contributes to  the $\gamma \Lambda \to \Lambda^{*}$ transition  while the isotriplet component gives contribution to the  $\gamma \Sigma^{0} \to \Lambda^{*}$ transition.  Thus, our result implies that the $\Lambda(1670)$ helicity amplitude has the isotriplet dominance, which is also found in  the nucleon helicity amplitude.

\begin{table}
\caption{Radiative decay width in units of keV.}
\label{tab:decaywidth}
\begin{center}
\begin{tabular}{cccc}
    \hline \hline
   &$\Lambda(1670)$ & $\Lambda(1390)$ & $\Lambda(1426)$ \\
   \hline
  $\Lambda^{*} \to \gamma \Lambda$    & $2.4$ & $19.2$ & $82.8$ \\
  $\Lambda^{*} \to \gamma \Sigma^{0}$ & $118.9$ & $112.9$ & $54.9$ \\
  \hline \hline
\end{tabular}
\end{center}
\end{table}

Using the result of the helicity amplitude and Eq.~(\ref{eq:decaywidth}), we calculate the radiative decay width of the $\Lambda(1670)$. The result is given in Table~\ref{tab:decaywidth}, reflecting a larger  transition amplitude for $\gamma \Sigma^{0} \to \Lambda^{*}$.  We find a large enhancement of the decay ratio of $\Gamma_{\gamma \Sigma^{0}}/\Gamma_{\gamma \Lambda}$, which is around 50. As previously discussed, this is a consequence of the presence of the strong $K\Xi$ channel. The same channel is also mainly responsible for the dynamical generation of the $\Lambda(1670)$, because its coupling constant to this channel is by far the largest one, as Table \ref{tab:coupl1680} shows. Thus, the large suppression of the $\Lambda^*\to\gamma \Lambda$ decay is directly tied to the nature of this resonance. This direct connection calls for an experimental test of the radiative decay widths.


\subsection{Results for the two \boldmath{$\Lambda(1405)$}}
\begin{figure}
\includegraphics[width=0.48\textwidth]{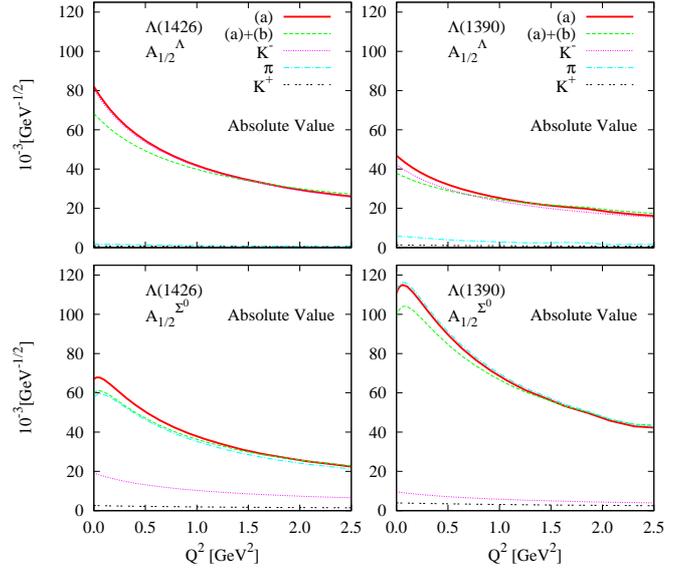} 
\caption{(Color Online) Absolute values of the $A_{1/2}^{p}$ helicity amplitudes for the $\Lambda(1390)$ (left) and $\Lambda(1426)$ (right) as a function of $Q^{2}$  in units of $10^{-3}$ GeV$^{-1/2}$.  The upper and lower  panels are for the $\gamma \Lambda \to \Lambda^{*}$ and  $\gamma \Sigma^{0} \to \Lambda^{*}$ transitions, respectively.  The solid lines show the calculation with diagram (a) (meson pole term).  The dashed line stands for the calculations of sum of diagrams (a) and (b).  The dotted, dot-dashed and two-dotted lines denote the $K^{-}$, pion and $K^{+}$  contributions, respectively.\label{fig:AbsLA}  }
\end{figure}

We also calculate the helicity amplitude for the two states of the $\Lambda(1405)$. For convenience, we call the $\Lambda(1405)$ states at the pole positions $z=1390-66i$ MeV and $z=1426-16i$ MeV as $\Lambda(1390)$ and $\Lambda(1426)$, respectively.  The calculation is done in the same formulation as the $\Lambda(1670)$ but using the coupling constants for $\Lambda(1390)$ and $\Lambda(1426)$ which are listed in Tables~\ref{tab:coupl1390} and \ref{tab:coupl1426}.

The results for the values of the helicity amplitudes at $Q^{2}=0$ are listed in Table~\ref{tab:A1/2value}, and the $Q^{2}$ dependence  of the helicity amplitudes is shown in Fig.~\ref{fig:AbsLA}. As already discussed in Ref.~\cite{mishageng}, the helicity amplitudes of the $\Lambda(1390)$ and $\Lambda(1420)$ can be understood by the coupling nature of the $\Lambda^{*}$. In Tables~\ref{tab:coupl1390} and \ref{tab:coupl1426}, the $\Lambda(1390)$ strongly couples to the $\pi\Sigma$ and $\bar KN$ states, while the $\Lambda(1426)$ couples to the $\bar KN$ state dominantly.  The couplings of these $\Lambda^{*}$ to $K \Xi$ are negligibly small.  As we discussed above, since in the $\gamma \Lambda \to \Lambda^{*}$ transition only the isosinglet photon component contributes to the  helicity amplitude in the isospin symmetric limit, the pion pole terms  give only a tiny contribution through the isospin breaking in the $\Lambda^{*}$ coupling constants. Thus, for both the $\Lambda(1390)$ and  $\Lambda(1426)$, the $K^{-}$ pole term gives the main contribution to the helicity amplitude of the $\gamma \Lambda \to \Lambda^{*}$ transition, and since the $\Lambda(1426)$ couples more strongly to  $K^{-}p$ than the $\Lambda(1390)$, the helicity amplitude of  the $\Lambda(1426)$ is larger than that of the $\Lambda(1390)$. 

For the $\gamma \Sigma^{0} \to \Lambda^{*}$ transition, in which  the isovector photon component contributes, the pion pole terms  are not suppressed by the symmetry argument, and the $\Sigma^{0}$ coupling to $\pi^{\pm}\Sigma^{\mp}$ is larger than that to $K^{-}p$. Thus, the pion pole terms are the main contributions to the  helicity amplitude for these two $\Lambda^{*}$.

In Table~\ref{tab:decaywidth}, we show the radiative decay widths of the  $\Lambda(1390)$ and $\Lambda(1420)$ calculated with the  helicity amplitude at $Q^{2}=0$. A detailed discussion on the comparison of the results for the radiative decays of these two states with data of the Particle Data Group was done in Ref.~\cite{mishageng} and we do not repeat it here. 


\subsection{Analytic continuation and uncertainties}
\label{sec:resana}
There are several uncertainties tied to the present results for the helicities amplitudes $A_{1/2}$ and radiative decays $\Gamma_{\gamma\Lambda}$, $\Gamma_{\gamma\Sigma^0}$. In the first place, we compare the radiative decay widths obtained at the energies of $z={\rm Re}\,z_0$ and $z=z_0$. In the previous sections, we have shown the results evaluated at $z={\rm Re}\,z_0$.	
In Sec.~\ref{sec:newana}, we have developed a scheme to analytically continue the loop functions. The radiative decay widths, evaluated using the analytic continuation, are given by $\Gamma_{\gamma\Lambda}=3.8$ keV and $\Gamma_{\gamma\Sigma^0}=119.6$ keV. As a comparison with Table \ref{tab:decaywidth} shows, these results are very similar for the $\gamma\Sigma^0$ decay but quite different for the $\gamma\Lambda$ decay. The reason is that the value for the $\gamma\Lambda$ decay involves a cancellation from large terms, as previously discussed. Then, any small change in these terms will lead to relatively large changes in the final result.
Note also that the poles of the $\Lambda(1670)$ and also the $\Lambda(1405)$ are relatively close to the physical axis at Re~$z=0$; for baryonic resonance with larger widths, one expects larger discrepancies when making the continuation to $z={\rm Re}\,z_0$, and in general, the evaluation at the pole position is preferable.

Second, to estimate the theoretical uncertainties, we have evaluated the radiative decay widths including also the subleading baryon pole diagram (b) from Fig.~\ref{fig:loops}. The radiative decay widths are given by $\Gamma_{\gamma\Lambda}=3.2$ keV and $\Gamma_{\gamma\Sigma^0}=71.4$ keV. This is a rather large change and shows that sub-leading diagrams can have a finite influence. As discussed in Ref.~\cite{Jido:2007sm}, there are other subleading terms from the normal and anomalous magnetic moments of the baryons and from the $\Lambda\Sigma^0$ transition magnetic moment, which can be of the same size as the subleading diagram (b) from Fig.~\ref{fig:loops}.

Third, we have evaluated the radiative decay widths using a different set of $D$ and $F$ constants, given by $D+F=1.26$ and $D-F=0.33$~\cite{mishageng} with the results $\Gamma_{\gamma\Lambda}=2.9$ keV and $\Gamma_{\gamma\Sigma^0}=104.7$ keV. 

Finally, uncertainties coming from the hadronic part of the amplitude, encoded in the decay constants $g_{\Lambda^*}$ from Eqs.~(\ref{eq:defA}) and (\ref{pa}), should be also estimated. In Ref.~\cite{bennhold}, where the $\Lambda(1670)$ was studied, the $K\Xi$ subtraction constant has been slightly adjusted to $a_{K\Xi}=-2.67$ with respect to the original value of $a_{K\Xi}=-2.52$ from Ref.~\cite{Jido:2003cb} ($a_{K\Xi}=-2.67$ is also the value used in this study; see Eq. (\ref{subtsorig}) for the set of used subtraction constants). This adjustment was justified because it left the properties of the two $\Lambda(1405)$ states almost untouched as their couplings to this channel is very small [cf. Tables \ref{tab:coupl1390} and \ref{tab:coupl1426}]. Using the original value of $a_{K\Xi}=-2.52$~\cite{Jido:2003cb}, the radiative decays widths of the $\Lambda(1670)$ are $\Gamma_{\gamma\Lambda}=1.14$ keV and $\Gamma_{\gamma\Sigma^0}=163$ keV.

Combining the theoretical uncertainties, we can assign final values of 
\be
\Gamma_{\gamma\Lambda}=3\pm 2\,\text{keV},\quad \Gamma_{\gamma\Sigma^0}=120\pm 50\,\text{keV}
\label{final}
\ee
to the radiative decay widths of the $\Lambda(1670)$.

Although the theoretical uncertainties are not small, the main point here is the pattern $\Gamma_{\gamma\Lambda}\ll \Gamma_{\gamma\Sigma^0}$ which appears as a stable feature when considering uncertainties. It is instructive to recall the situation of the two radiative decay widths of the $\Lambda(1520)$, calculated in Ref.~\cite{mishasourav}. There, the theoretical prediction delivers $\Gamma_{\gamma\Lambda}\ll \Gamma_{\gamma\Sigma^0}$ like in the present study, but the available experimental data for the $\gamma\Lambda$ decay is in disagreement with the $\Gamma_{\gamma\Lambda}$ decay~\cite{Taylor:2005zw} (the decay into $\gamma\Sigma^0$ is in agreement). The conclusion was that some genuine 3-quark component for the $\Lambda(1520)$ is needed; thus, experimental information for the present case of the $\Lambda(1670)$ would be useful to further shed light on the nature of this resonance.  

It is interesting to compare the present values to predictions from the quark model of Ref.~\cite{VanCauteren:2005sm}. There, the radiative decays are given by $\Gamma_{\gamma\Lambda}=0.159$ keV and $\Gamma_{\gamma\Sigma^0}=3827$ keV. While the same pattern $\Gamma_{\gamma\Lambda} \ll \Gamma_{\gamma\Sigma^0}$ is observed in this quark model, the absolute size of $\Gamma_{\gamma\Sigma^0}$ is much larger than the present prediction from the chiral unitary framework. An experiment that measures both radiative decays would thus be most welcome to decide between both scenarios or point towards new mechanisms considered in none of the theoretical approaches.


\section{Conclusion}
We have determined the helicity amplitudes and form factor for 
the $A_{1/2}$ amplitudes for the decay of the $\Lambda(1670)$ and 
the two $\Lambda(1405)$ into $\gamma \Lambda$ and $\gamma \Sigma$.
We find quite different results in all cases. Particularly striking is the case 
of the $\Lambda(1670)$ decay into $\gamma \Lambda$ and 
$\gamma \Sigma$, where the amplitudes differ by one 
order of magnitude. We could trace back these results to the peculiar 
structure of the resonances as dynamically generated from the 
interaction of the pseudoscalar meson octet and the baryon octet, since 
there are large cancellations from the contributions of the different 
meson baryon components in the case of the $\Lambda(1670)$ decay into 
$\gamma \Lambda$. This picture is quite different from the one of the quark 
models as a consequence of which we also find big differences with the 
results of relativistic quark models. No doubt, measurements of these 
magnitudes would be very valuable in the quest for determining the 
nature of these resonances. 
We have also developed a technique to evaluate the helicity amplitudes
at the resonance pole position by making an analytic continuation of the 
Feynman parameterized integrals to the second Riemann sheet. 

\begin{acknowledgement}
D.J. wishes to acknowledge the hospitality of the University of Valencia, where part of this work was done. The work of M. D. is supported by the DFG (Deut\-sche For\-schungs\-gemeinschaft, Gz: DO 1302/1-2).  This work is partly supported by DGICYT Contract No. FIS2006-03438, the Generalitat Valenciana in the program Prometeo and the EU Integrated Infrastructure Initiative Hadron Physics Project under contract RII3-CT-2004-506078 and the DFG under contract No. GRK683. We also acknowledge the collaboration agreement between the JSPS of Japan and the CSIC of Spain, the Grant for Scientific Research (No.\ 20028004) and the Grant-in-Aid for the Global COE Program ``The Next Generation of Physics, Spun from Universality and Emergence" from the Ministry of Education, Culture, Sports, Science and Technology (MEXT) of Japan. 
This research 
was done under Yukawa International Program for Quark-Hadron Sciences. We would like to thank U.-G. Mei{\ss}ner for a careful reading of the manuscript.
\end{acknowledgement}


\appendix
\section{Model of \boldmath$\Lambda(1670)$}
\label{app:L1670}

In this appendix, the model for the $\Lambda(1670)$ is briefly reviewed.  In the chiral unitary approach, the $\Lambda(1670)$ resonance is  dynamically generated in $s$-wave meson baryon scattering  in the coupled channels of $K^{-}p$, $\bar K^{0}n$, $\pi^{0}\Lambda$, $\pi^{0}\Sigma^{0}$, $\eta\Lambda$, $\eta \Sigma^{0}$, $\pi^{+}\Sigma^{-}$, $\pi^{-}\Sigma^{+}$, $K^{+}\Xi^{-}$ and $K^{0} \Xi^{0}$ without introducing an explicit pole term in the fundamental interaction. 

The scattering amplitude for the $\Lambda(1670)$ resonance was described  in Refs. \cite{joseulf,bennhold} with the scattering equation for the meson  baryon system given by
\begin{equation}
   T = V + V G T \ .
\end{equation}
Based on the $N/D$ method and the dispersion relation \cite{joseulf}, this integral scattering equation can be reduced to a simple algebraic equation
\begin{equation}
   T = (1-VG)^{-1}\,V
   \label{bse}
\end{equation}
where the matrix $V$ is the interaction kernel of the scattering equation and we take  the $s$-wave meson-baryon interaction given by the lowest order of the chiral  perturbation theory (the Weinberg-Tomozawa interaction), which is given by
\begin{eqnarray}
V_{i j} (W) &=&
 - C_{i j} \frac{1}{4 f^2}(2W - M_{i}-M_{j})
 \nonumber \\ &&  \ \ \ \times
\sqrt{\frac{M_{i}+E_{i}}{2M_{i}}}
\sqrt{\frac{M_{j}+E_{j}}{2M_{j}}}
\label{eq:ampl2}
\end{eqnarray}
with the channel indices $i,j$, the baryon mass $M$, the meson mass $m$,  the meson decay constant $f$, the baryon energy $E$ and the center of mass energy $W$ in the meson-baryon system. The coefficient $C_{ij}$ is the coupling strength of the meson  and baryon, which is determined solely by the flavor SU(3) structure of the channel.  The diagonal matrix $G$ in the flavor space is the meson baryon loop function  given in terms of the meson and baryon propagators by
\begin{eqnarray}
   G(W) &=&  i\int \frac{d^{4} q}{(2\pi)^{4}} \frac{M}{E(\vec q)}
    \frac{1}{q^{0} -E(\vec q) + i\epsilon}
   \nonumber \\ && \ \ \ \ \ \ \times
    \frac{1}{(P-q)^{2} - m^{2} + i \epsilon}
    \label{propnorela}
\end{eqnarray}
with  the total energy $P=(W, 0, 0,0)$ in the center of mass frame.  For the baryon propagator we have used the nonrelativistic form and   neglected the negative energy propagation. In the practical calculation, we use  a covariant form of the positive energy part of the baryon propagator,
\begin{eqnarray}
\frac{M}{E(\vec q\, )}\,\frac{\Sigma_r u_r(\vec q\, )\overline{u}_r(\vec q\, )}{q^0-E(\vec q\,)+i\epsilon}\simeq
\frac{2M \,\Sigma_r u_r(\vec q\, )\overline{u}_r(\vec q\, )}{q^2-M^2+i\epsilon},
\label{covaform}
\end{eqnarray}
where $u(\vec q)$ is the Dirac spinor for the baryon and we sum up in terms of the spin index $r$.

The divergent loop function should be regularized with proper schemes.  We take dimensional regularization, since it is equivalent to the dispersion  integral of the two-body meson-baryon phase space with one subtraction  and, therefore, we can perform the integral  in a consistent way with analyticity of the loop function in terms of  the energy $W$~\cite{joseulf}. In dimensional regularization,  the loop function in each channel $i$ is  given by the following analytic expression:   
\begin{eqnarray}
G_i(W)
&=& i \,  \int \frac{d^4 q}{(2 \pi)^4} \,
\frac{2 M_i}{q^2 - M_i^2 + i \epsilon} \, \frac{1}{(P-q)^2 - m_i^2 + i
\epsilon}\non
 &=& \frac{2 M_i}{16 \pi^2} \left\{ a_i(\mu) + \ln
\frac{M_i^2}{\mu^2} + \frac{m_i^2-M_i^2 + W^{2}}{2W^{2}} \ln \frac{m_i^2}{M_i^2} 
\right. \nonumber \\ &  +&
\frac{\bar q_i}{W}
\left[
\ln(\hspace*{0.2cm}W^{2}-(M_i^2-m_i^2)+2 \bar q_i W)
\right. 
 \non 
&&\hspace*{0.33cm} + \ln(\hspace*{0.2cm}W^{2}+(M_i^2-m_i^2)+2 \bar q_i W) \non
&&\hspace*{0.33cm}-  \ln(-W^{2}+(M_i^2-m_i^2)+2\bar q_i W) \non 
&&\left. \left. \hspace*{0.2cm}  - \ln(-W^{2}-(M_i^2-m_i^2)+2\bar q_i W) \right]
\right\} ,
\label{eq:gpropdr}
\end{eqnarray}
where $\bar q_i\equiv q_{cm}$ with $q_{cm}$ from Eq. (\ref{limits}) is the 3-momentum of the meson or baryon in the center of mass frame and  $\mu$ is the scale of dimensional regularization. The remaining finite constant terms denoted by $a_i(\mu)$ are determined  phenomenologically by a fit so as to reproduce the threshold branching ratios of $K^{-}p$ to $\pi\Lambda$ and $\pi\Sigma$ observed  by stopped $K^{-}$ mesons in hydrogen~\cite{Tovee:1971ga,Nowak:1978au}, as done  in Refs. \cite{Kaiser:1995eg,joseulf,bennhold}. The values of these constants are very important for the nature of the dynamically  generated resonances. It was pointed out in Ref.~\cite{Hyodo:2008xr} that, although we consider only the meson-baryon dynamics in the chiral unitary approach, depending on the values of the $a_{i}$ constants, some other components than meson and baryon in the formulation, such as genuine quark components, can be involved implicitly into the scattering amplitude.  Here we use the following $a_{i}$ constants determined in Ref.~\cite{bennhold}:
\begin{equation}
  \begin{array}{ccc}
   a_{\bar K N} = -1.84, \hspace{0.3cm}& a_{\pi \Sigma} = - 2.00,
   \hspace{0.3cm}& a_{\pi \Lambda} = -1.83 \\
   a_{\eta \Lambda} = -2.25, \hspace{0.3cm} & a_{\eta \Sigma} = -2.38,
   \hspace{0.3cm} & a_{K \Xi} = -2.67 
   \end{array}
   \label{subtsorig}
\end{equation}
with $\mu =630$ MeV. These values are close to a natural value equivalent to the three-momentum cut-off with 630 MeV/c~\cite{joseulf} and it turned out that with these values the $\Lambda(1405)$ is described predominantly  by meson-baryon components~\cite{Hyodo:2008xr}. For the other parameters, we use $f=1.123 f_{\pi}$  with $f_{\pi}=93$ MeV and the observed meson and baryon masses (small  isospin breaking effects come also into the scattering amplitude due to the  isospin breaking from the physical meson and baryon masses). Once these constants are fixed, the amplitudes involving photons can be  predicted without introducing any new free parameters.

The amplitudes $T^{ij}$ obtained from Eq.~(\ref{bse}) can be analytically continued  to the complex plane of the scattering energy $W$.  The amplitude has poles in the complex plane that are identified with the resonances. The residua of the poles determine the coupling strengths $g_{\Lambda^{*}}^{i}$  of the resonances to the meson-baryon channels. Thus, the scattering amplitude  for the channels $i$ and $j$ close to the resonance is written as given in Eq. (\ref{pa}).
The pole position for the $\Lambda(1670)$ resonance is obtained~\cite{bennhold} as 
\begin{equation}
  z = 1680 -20i  \ \ [{\rm MeV}] \ . \label{eq:pole1680}
\end{equation}
If we use the natural renormalization scheme suggested in Ref.~\cite{Hyodo:2008xr} which excludes the CDD pole component, we obtain a pole at $1700-21i$ MeV. This pole  is not so far from the pole of Eq. (\ref{eq:pole1680}), which is obtained with the  parameters of Eq. (\ref{subtsorig}) determined phenomenologically by fitting the threshold branching ratio. This implies that the $\Lambda(1670)$ is also described almost by the meson-baryon dynamics.  In Table \ref{tab:coupl1680} we list the values of the coupling constants $g_{\Lambda^{*}}^{i}$, which characterize the structure of the $\Lambda(1670)$. The $\Lambda(1670)$ has  large couplings to the $\eta \Lambda$ and $K \Xi$ channels. Especially $K^{+}\Xi^{-}$ is a relevant channel for the photon coupling. 

The scattering amplitudes obtained here also include the $\Lambda(1405)$. For the $\Lambda(1405)$ it is known from the investigations of  Refs.~\cite{joseulf,Jido:2002yz,Jido:2003cb} that there are two poles at
\begin{equation}
   z = 1390 -66i \ \ [{\rm MeV}] \label{eq:pole1390}
\end{equation}
and
\begin{equation}
   z = 1426 -16i \ \ [{\rm MeV}]  \ .  \label{eq:pole1426}
\end{equation}
Here we call these two poles for the $\Lambda(1405)$ as $\Lambda(1390)$ and  $\Lambda(1426)$ for convenience. The couplings of these poles are listed in  Tables~\ref{tab:coupl1390} and \ref{tab:coupl1426}. Experimental support for the double pole structure of the $\Lambda(1405)$ is found in Refs.~\cite{Hyodo:2004vt,Magas:2005vu,Jido:2009jf}.

\begin{table*}
\caption{Coupling constants {$g_i$} of $\Lambda(1390)$ to the meson-baryon channels obtained as the residua of the scattering amplitude at the pole position $z=1390 - 66i$ MeV.}
\begin{center}
\begin{tabular}{cccccc}
 \hline\hline
channel &   $pK^{-} $ & $n \bar K^{0}$ & $\Lambda \pi^{0}$ & $\Sigma^{0} \pi^{0}$  & $\Lambda \eta$ \\
$g_{\Lambda^{*}}^{i}$ &   $ -0.889 +  1.232 i $ & $-0.813 + 1.226 i $ & $ -0.005 + 0.013 i$ & $-1.418 + 0.877 i$ & $-0.006 + 0.759 i$ \\ 
$|g_{\Lambda^{*}}^{i}|$&   $1.520$ & $1.471$ & $0.014$ & $1.667$ & $0.759$ \\
 \hline
channel &$\Sigma^{0} \eta$ &$\Sigma^{-} \pi^{+} $& $\Sigma^{+} \pi^{-}$ & $\Xi^{-} K^{+}$ & $\Xi^{0} K^{0}$ \\
$g_{\Lambda^{*}}^{i}$ &  $  -0.004 + 0.011 i $& $-1.358 + 0.868 i$ & $  -1.477 + 0.870 i$  & $ -0.302 + 0.292 i$ & $-0.340 +  0.288 i$\\
$|g_{\Lambda^{*}}^{i}|$&    $0.012$ & $1.612$ & $1.715$ & $0.420$ & $0.446$ \\
  \hline\hline
\end{tabular}
\end{center}
\label{tab:coupl1390}
\end{table*}
\begin{table*}
\caption{Coupling constants {$g_i$} of $\Lambda(1426)$ to the meson-baryon channels obtained as the residua of the scattering amplitude at the pole position $z=1426 - 16i$ MeV.}
\begin{center}
\begin{tabular}{cccccc}
 \hline\hline
channel &
   $pK^{-} $ & $n \bar K^{0}$ & $\Lambda \pi^{0}$ & $\Sigma^{0} \pi^{0}$  & $\Lambda \eta$ \\
$g_{\Lambda^{*}}^{i}$ &   $1.839 + 0.689 i $ & $1.759 + 0.646 i $ & $0.042 -0.003 i$ & $ 0.242 + 0.839 i$ & $1.396 +  0.208 i$ \\ 
$|g_{\Lambda^{*}}^{i}|$&   $1.964$ & $1.874$ & $0.042$ & $0.873$ & $1.411$ \\
 \hline
channel &$\Sigma^{0} \eta$ &$\Sigma^{-} \pi^{+} $& $\Sigma^{+} \pi^{-}$ & $\Xi^{-} K^{+}$ & $\Xi^{0} K^{0}$ \\
$g_{\Lambda^{*}}^{i}$ &  $ 0.037 -0.003 i$& $0.222 + 0.814 i$ & $  0.255 + 0.863 i$  & $0.086 + 0.228 i$ & $0.070 +  0.238 i$\\
$|g_{\Lambda^{*}}^{i}|$&    $0.037$ & $0.843$ & $0.900$ & $0.243$ & $0.248$ \\
  \hline\hline
\end{tabular}
\end{center}
\label{tab:coupl1426}
\end{table*}


\begin{thebibliography}{00}
\bibitem{Isgur:1978xj}
  N.~Isgur and G.~Karl,
  Phys.\ Rev.\  D {\bf 18}, 4187 (1978).
\bibitem{Capstick:1986bm}
  S.~Capstick and N.~Isgur,
  Phys.\ Rev.\  D {\bf 34}, 2809 (1986).
\bibitem{Glozman:1995fu}
  L.~Y.~Glozman and D.~O.~Riska,
  Phys.\ Rept.\  {\bf 268}, 263 (1996).
\bibitem{Glozman:1996wq}
  L.~Y.~Glozman, Z.~Papp and W.~Plessas,
  Phys.\ Lett.\  B {\bf 381}, 311 (1996).
\bibitem{Capstick:2000qj}
  S.~Capstick and W.~Roberts,
  Prog.\ Part.\ Nucl.\ Phys.\  {\bf 45}, S241 (2000).
\bibitem{Loring:2001kv}
  U.~L\"oring, K.~Kretzschmar, B.~C.~Metsch and H.~R.~Petry,
  Eur.\ Phys.\ J.\  A {\bf 10}, 309 (2001).
\bibitem{Merten:2002nz}
  D.~Merten, U.~L\"oring, K.~Kretzschmar, B.~Metsch and H.~R.~Petry,
  Eur.\ Phys.\ J.\  A {\bf 14}, 477 (2002).
\bibitem{Furuichi:2003eh}
  M.~Furuichi, K.~Shimizu and S.~Takeuchi,
  Phys.\ Rev.\  C {\bf 68}, 034001 (2003).
\bibitem{VanCauteren:2003hn}
  T.~Van Cauteren, D.~Merten, T.~Corthals, S.~Janssen, B.~Metsch, H.~R.~Petry and J.~Ryckebusch,
  Eur.\ Phys.\ J.\  A {\bf 20}, 283 (2004).
\bibitem{VanCauteren:2005sm}
T.~Van Cauteren, J.~Ryckebusch, B.~Metsch and H.-R.~Petry,
Eur. Phys. J. {\bf A26} (2005) 339.


\bibitem{Drechsel:1998hk}
  D.~Drechsel, O.~Hanstein, S.~S.~Kamalov and L.~Tiator,
  Nucl.\ Phys.\  A {\bf 645} (1999) 145.

 \bibitem{Penner:2002ma}
  G.~Penner and U.~Mosel,
  Phys.\ Rev.\  C {\bf 66}, 055211 (2002).

\bibitem{Gasparyan:2003fp}
  A.~M.~Gasparyan, J.~Haidenbauer, C.~Hanhart and J.~Speth,
  Phys.\ Rev.\  C {\bf 68}, 045207 (2003).

\bibitem{Matsuyama:2006rp}
  A.~Matsuyama, T.~Sato and T.~S.~Lee,
  Phys.\ Rept.\  {\bf 439} (2007) 193.

\bibitem{Durand:2008es}
  J.~Durand, B.~Juli\'a-D\'iaz, T.~S.~Lee, B.~Saghai and T.~Sato,
  Phys.\ Rev.\  C {\bf 78}, 025204 (2008).

\bibitem{Doring:2009yv}
  M.~D\"oring, C.~Hanhart, F.~Huang, S.~Krewald and U.-G.~Mei{\ss}ner,
  Nucl. \ Phys.\ A {\bf 829}, 170 (2009).


\bibitem{Kaiser:1995eg}
N. Kaiser, P. B. Siegel and W. Weise,
Nucl. Phys. {\bf A594} (1995) 325.


\bibitem{kaiser}
  N.~Kaiser, T.~Waas and W.~Weise,
  Nucl.\ Phys.\ A {\bf 612}, 297 (1997).

\bibitem{angels}
  E.~Oset and A.~Ramos,
  Nucl.\ Phys.\ A {\bf 635}, 99 (1998).

\bibitem{joseulf}
  J.~A.~Oller and U.-G.~Mei\ss ner,
  Phys.\ Lett.\ B {\bf 500}, 263 (2001).

\bibitem{bennhold}
  E.~Oset, A.~Ramos and C.~Bennhold,
  Phys.\ Lett.\ B {\bf 527}, 99 (2002)
  [Erratum-ibid.\ B {\bf 530}, 260 (2002)].

\bibitem{Jido:2003cb}
  D.~Jido, J.~A.~Oller, E.~Oset, A.~Ramos and U.-G.~Mei\ss ner,
  Nucl.\ Phys.\ A {\bf 725}, 181 (2003).

\bibitem{nieves}
  C.~Garcia-Recio, J.~Nieves, E.~Ruiz Arriola and M.~J.~Vicente Vacas,
  Phys.\ Rev.\ D {\bf 67}, 076009 (2003).

\bibitem{carmen}
  C.~Garcia-Recio, M.~F.~M.~Lutz and J.~Nieves,
  Phys.\ Lett.\ B {\bf 582}, 49 (2004).

\bibitem{hyodo}
  T.~Hyodo, S.~I.~Nam, D.~Jido and A.~Hosaka,
  Phys.\ Rev.\ C {\bf 68}, 018201 (2003);
  Prog.\ Theor.\ Phys.\  {\bf 112} (2004) 73.
  
\bibitem{lutz}
  E.~E.~Kolomeitsev and M.~F.~M.~Lutz,
  Phys.\ Lett.\  B {\bf 585} (2004) 243.
  
\bibitem{sarkar}
  S.~Sarkar, E.~Oset and M.~J.~Vicente Vacas,
  Nucl.\ Phys.\  A {\bf 750} (2005) 294
  [Erratum-ibid.\  A {\bf 780} (2006) 78].

\bibitem{Borasoy:2006sr}
  B.~Borasoy, U.-G.~Mei{\ss}ner and R.~Ni{\ss}ler,
  Phys.\ Rev.\  C {\bf 74}, 055201 (2006).
  
\bibitem{Oller:2006jw}
  J.~A.~Oller,
  Eur.\ Phys.\ J.\  A {\bf 28}, 63 (2006).


\bibitem{Hyodo:2008xr}
  T.~Hyodo, D.~Jido and A.~Hosaka,
  Phys.\ Rev.\  C {\bf 78} (2008) 025203.

\bibitem{Dalitz:1959dn}
  R.~H.~Dalitz and S.~F.~Tuan,
  Phys.\ Rev.\ Lett.\  {\bf 2}, 425 (1959).

\bibitem{Dalitz:1960du}
  R.~H.~Dalitz and S.~F.~Tuan,
  Annals Phys.\  {\bf 10}, 307 (1960).


\bibitem{Jido:2007sm}
  D.~Jido, M.~D\"oring and E.~Oset,
  Phys.\ Rev.\  C {\bf 77} (2008) 065207.

\bibitem{Jido:2008fr}
  D.~Jido, T.~Hyodo and A.~Hosaka,
  Mod.\ Phys.\ Lett.\  A {\bf 23} (2008) 2389.

\bibitem{Doring:2009qr}
  M.~D\"oring and K.~Nakayama,
  Eur. Phys. J. A {\bf 43}, 83 (2010).

\bibitem{mishasourav}
  M.~D\"oring, E.~Oset and S.~Sarkar,
  Phys.\ Rev.\  C {\bf 74} (2006) 065204.

\bibitem{Nacher:1999ni}
  J.~C.~Nacher, E.~Oset, H.~Toki and A.~Ramos,
  Phys.\ Lett.\  B {\bf 461}, 299 (1999).

\bibitem{Jido:2002yz}
  D.~Jido, A.~Hosaka, J.~C.~Nacher, E.~Oset and A.~Ramos,
  Phys.\ Rev.\  C {\bf 66} (2002) 025203.

\bibitem{mishasolo}
  M.~D\"oring,
  Nucl.\ Phys.\  A {\bf 786}, 164 (2007).
  
\bibitem{mishageng}
  L.~S.~Geng, E.~Oset and M.~D\"oring,
  Eur.\ Phys.\ J.\  A {\bf 32}, 201 (2007).
  
\bibitem{Sekihara:2008qk}
  T.~Sekihara, T.~Hyodo and D.~Jido,
  Phys.\ Lett.\  B {\bf 669} (2008) 133.

\bibitem{Ramos:2003wt}
  A.~Ramos, E.~Oset, C.~Bennhold, D.~Jido, J.~A.~Oller and U.-G.~Mei{\ss}ner,
  Nucl.\ Phys.\  A {\bf 754}, 202 (2005).

\bibitem{Borasoy:2005zg}
  B.~Borasoy, P.~C.~Bruns, U.-G.~Mei{\ss}ner and R.~Ni{\ss}ler,
  Phys.\ Rev.\  C {\bf 72}, 065201 (2005).

\bibitem{Borasoy:2007ku}
  B.~Borasoy, P.~C.~Bruns, U.-G.~Mei{\ss}ner and R.~Ni{\ss}ler,
  Eur. Phys. J. A {\bf 34} (2007), 161.

\bibitem{Doring:2009bi}
  M.~D\"oring, C.~Hanhart, F.~Huang, S.~Krewald and U.-G.~Mei{\ss}ner,
  Phys.\ Lett.\  B {\bf 681}, 26 (2009).
  
\bibitem{Drechsel:2007if}
  D.~Drechsel, S.~S.~Kamalov and L.~Tiator,
  Eur.\ Phys.\ J.\  A {\bf 34}, 69 (2007).

  
\bibitem{Aznauryan:2009mx}
  I.~G.~Aznauryan {\it et al.}  [CLAS Collaboration],
  Phys.\ Rev.\  C {\bf 80}, 055203 (2009).

\bibitem{Doring:2009uc}
  M.~D\"oring and K.~Nakayama,
  Eur.\ Phys.\ J.\  A {\bf 43}, 83 (2010).

\bibitem{Luty:1993gi}
  M.~A.~Luty and M.~J.~White,
  Phys.\ Lett.\  B {\bf 319} (1993) 261.
\bibitem{Bernard:1992nc}
  V.~Bernard, N.~Kaiser and U.-G.~Mei{\ss}ner,
  Nucl.\ Phys.\  B {\bf 383}, 442 (1992).
\bibitem{Bernard:1994gm}
  V.~Bernard, N.~Kaiser and U.-G.~Mei{\ss}ner,
  Z.\ Phys.\  C {\bf 70}, 483 (1996).

\bibitem{Amendolia:1986wj}
  S.~R.~Amendolia {\it et al.}  [NA7 Collaboration],
  Nucl.\ Phys.\  B {\bf 277}, 168 (1986).

\bibitem{Amendolia:1986ui}
  S.~R.~Amendolia {\it et al.},
  Phys.\ Lett.\  B {\bf 178}, 435 (1986).
  
\bibitem{Oller:2000ug}
  J.~A.~Oller, E.~Oset and J.~E.~Palomar,
  Phys.\ Rev.\  D {\bf 63} (2001) 114009.

\bibitem{Suzuki:2008rp}
  N.~Suzuki, T.~Sato and T.~S.~Lee,
  Phys.\ Rev.\  C {\bf 79}, 025205 (2009).

\bibitem{Taylor:2005zw}
  S.~Taylor {\it et al.}  [CLAS Collaboration],
  Phys.\ Rev.\  C {\bf 71}, 054609 (2005)
  [Erratum-ibid.\  C {\bf 72}, 039902 (2005)].

\bibitem{Tovee:1971ga}
D. N. Tovee {\it et al.},
Nucl. Phys. {\bf B33} (1971) 493.

\bibitem{Nowak:1978au}
R. J. Nowak {\it et al.},
Nucl. Phys. {\bf B139} (1978) 61.

\bibitem{Hyodo:2004vt}
  T.~Hyodo, A.~Hosaka, M.~J.~Vicente Vacas and E.~Oset,
  Phys.\ Lett.\  B {\bf 593}, 75 (2004).

\bibitem{Magas:2005vu}
  V.~K.~Magas, E.~Oset and A.~Ramos,
  Phys.\ Rev.\ Lett.\  {\bf 95}, 052301 (2005).

\bibitem{Jido:2009jf}
  D.~Jido, E.~Oset and T.~Sekihara,
  Eur.\ Phys.\ J.\  A {\bf 42}, 257 (2009).


\end{thebibliography}
\end{document}